\documentclass[%
 reprint,
nofootinbib,
 amsmath,amssymb,
 aps
]{revtex4-1}
\pdfoutput=1
\usepackage{graphicx}
\usepackage[utf8]{inputenc}
\usepackage{flushend}
\usepackage{dcolumn}
\usepackage{bm}
\usepackage{balance}

\usepackage[colorlinks = true,
            linkcolor = blue,
            urlcolor  = blue,
            citecolor =red,
            anchorcolor = blue]{hyperref}
\usepackage{verbatim}
\usepackage{color,ulem}
\usepackage[english]{babel}

\usepackage[utf8]{inputenc}

\begin{document}

\preprint{}{IFT-UAM/CSIC-19-126, IPM/P-2019/039}

\title{\huge Nonlinear Oscillatory Shear Tests \\ in Viscoelastic Holography}

\author{Matteo Baggioli}%
 \email{matteo.baggioli@uam.es}
\affiliation{Instituto de Fisica Teorica UAM/CSIC, c/Nicolas Cabrera 13-15,
Universidad Autonoma de Madrid, Cantoblanco, 28049 Madrid, Spain.
}%

\author{Sebastian Grieninger}%
 \email{sebastian.grieninger@gmail.com}
\affiliation{Theoretisch-Physikalisches Institut, Friedrich-Schiller-Universit\"at Jena,
Max-Wien-Platz 1, D-07743 Jena, Germany.
}%
\affiliation{Department of Physics, University of Washington, Seattle, WA 98195-1560, USA}

\author{Hesam Soltanpanahi}%
 \email{hesam@m.scnu.edu.cn}
\affiliation{Institute of Quantum Matter, School of Physics and Telecommunication Engineering, South China
Normal University, Guangzhou 510006, China ;
}%

\affiliation{School of Physics, Institute for Research in Fundamental Sciences (IPM), P.O.Box 19395-5531,
Teheran, Iran}

\affiliation{M. Smoluchowski Institute of Physics and Mark Kac Center for Complex Systems Research, Jagiellonian University, 30-348 Krakw, Poland.}

\begin{abstract}
We provide the first characterization of the nonlinear  and  time  dependent  rheologic response of viscoelastic bottom-up holographic models. More precisely, we perform oscillatory shear tests in holographic massive gravity theories with finite elastic response, focusing on the large amplitude oscillatory shear (LAOS) regime. The characterization of these systems is done using several techniques: (I) the Lissajous figures, (II) the Fourier analysis of the stress signal, (III) the Pipkin diagram and (IV) the dependence of the storage and loss moduli on the amplitude of the applied strain. We find substantial evidence for a strong \textit{strain stiffening} mechanism, typical of hyper-elastic materials such as rubbers and complex polymers. This indicates that the holographic models considered are not a good description for rigid metals, where strain stiffening is not commonly observed.  Additionally, a crossover between a \textit{viscoelastic liquid} regime at small graviton mass (compared to the temperature scale), and a \textit{viscoelastic solid} regime at large values is observed. Finally, we discuss the relevance of our results for soft matter and for the understanding of the widely used homogeneous holographic models with broken translations.

\end{abstract}

\maketitle

\section{Introduction} 
In elastic solids, the mechanical stress is proportional to the applied external shear strain \cite{landau1986theory}. However, in hydrodynamic fluids the stress is proportional to the shear rate \cite{landau2013fluid}. Of course, both of these cases are abstract idealizations, valid only under limiting conditions. In general, all the materials are \textit{viscoelastic} -- they present an interplay between elastic effects and dissipative viscous ones \cite{christensen1971theory}; honey is the most common example.\\
The idea that ''\textit{everything flows if you wait long enough}'' lies behind the foundation of a new field of research known as \textit{rheology} \cite{barnes1989introduction,larson1999structure} -- the study of deformation and flow of matter. Even though the first model goes back to Maxwell in 1867 \cite{Roylance01engineeringviscoelasticity}, a large part of the theoretical description of viscoelastic materials is still based on phenomenological frameworks (Kelvin-Voigt, generalized Maxwell, Burgers)~\cite{malkin2006rheology}.\\
The fundamental difficulties are twofold: (I) it is conceptually hard to incorporate dissipation into the effective field theory description of solid materials \cite{nicolis2015zoology,Endlich:2012vt} because of the unavoidable requirement of unitarity; (II)  elasticity theory can be formulated in the language of standard effective field theory following a well defined action principle \cite{Nicolis:2015sra}. Hydrodynamics, on the contrary, is usually described by a set of conservation equations and constitutive relations \cite{Kovtun:2012rj} and it is not suitable for a description in terms of a local and hermitian action \cite{Glorioso:2018wxw}. \\
The problem becomes even more acute when the amplitude of the applied external strain is not small and the linear approximation is of no help anymore -- the onset of nonlinear viscoelasticity \cite{lockett1972nonlinear,Holzapfel2002,wang2018nonlinear,pipkin1986lectures}.
For simplicity, in this letter, we focus on oscillatory shear tests in which the external shear strain takes a simple sinusoidal form 
\begin{equation}
 \gamma(t)\,=\,\gamma_0\,\sin(2 \pi\omega t)
\end{equation}
where $\gamma_0$ is the strain amplitude and $\omega$ is the characteristic frequency.
A convenient characterization of these rheology experiments is defined through the Deborah number $\textit{De}=\omega\,\lambda$ and the Weissenberg number $\textit{Wi}\,\equiv\,\lambda\,\gamma_0$, where $\lambda$ is the characteristic relaxation time of the material -- in simple words, these two numbers determine how fast and how strong we are probing the viscoelastic system.
Small \textit{Wi} and large \textit{De} corresponds to linear elasticity, in which the stress output is linearly dependent on the external input.\\
Generally speaking, we can draw the so-called Pipkin diagram \cite{pipkin1986lectures}, picturing the phase space of the system in function of the values of these two numbers. In the largest region of this diagram, strain amplitudes are large and frequencies are neither high nor low; experiments probing that region are called \textit{LAOS tests} \cite{HYUN20111697,4f5dcea3a806435da7b4b4cbadb06d63} and are the subject of this letter.\\
In the LAOS regime, linear viscoelasticity is not applicable anymore; the response is fully nonlinear, the storage and loss moduli become non-trivial functions of the strain amplitude $\gamma_0$. Very little is known in this regime, quoting Pipkin himself: ''\textit{Here Be Dragons}'' \cite{pipkin1986lectures}.\\
From a totally different perspective, in the last ten years, Holography revealed to be a very useful tool for the development and understanding of Hydrodynamics~\cite{Policastro:2002se,Policastro:2002tn,Janik:2006ft,Kovtun:2005ev,Kovtun:2004de,Kovtun:2003wp}. The most famous examples are: (I) the formulation of a universal bound on the viscosity-to-entropy ratio \cite{Policastro:2001yc} which is so far respected by all known fluids \cite{Cremonini:2011iq}; (II) the discovery of new transport coefficients in anomalous hydrodynamics \cite{Landsteiner:2011cp,Landsteiner:2011iq} experimentally observed in Weyl semimetals \cite{Gooth:2017mbd}.\\
A fundamental breakthrough in this direction is the observation that black holes (BHs) behave as dissipative hydrodynamic systems \cite{1986bhmp.book.....T,Iqbal:2008by}. From this point of view, the application of an external strain source to the hydrodynamic system corresponds to the perturbation of the BH geometry by dynamical gravitational waves \cite{Policastro:2002se}. \\
More recently, a series of works \cite{Baggioli:2014roa,Alberte:2015isw,Alberte:2017cch,Alberte:2017oqx} explained how to endow black holes with a solid structure, providing them with a finite elastic response. In these new holographic theories, the BH response is no longer purely hydrodynamic but it becomes viscoelastic \cite{Andrade:2019zey} in all aspects. Since then, a lot of effort has been devoted to the implementation, the classification and the characterization of these setups and similar ones \cite{Alberte:2016xja,Baggioli:2019abx,Ammon:2019apj,Ammon:2019wci,Baggioli:2018nnp,Baggioli:2018vfc,Baggioli:2018bfa,Grozdanov:2018ewh,Amoretti:2017frz,Armas:2019sbe,Esposito:2017qpj,Fukuma:2012ws}.
In this letter, we provide the first characterization of the nonlinear and time dependent viscoelastic response of these holographic models, with particular emphasis on the LAOS regime. The relevance of our results is diverse and highly interdisciplinary: (I) to shed light on the challenge of LAOS and in particular the physics of complex fluids (yelding, shear thinning, stress overshoot, dynamical instabilities) \cite{4f5dcea3a806435da7b4b4cbadb06d63};
(II) to reach a full characterization and understanding of the homogeneous holographic models with broken translations \cite{Baggioli:2014roa,Andrade:2013gsa,Donos:2013eha,Vegh:2013sk} and their possible connections with glasses, complex fluids and amorphous systems \cite{Facoetti:2019rab}; (III) to study out of equilibrium processes in strongly coupled field theories and the possible universal evolution after dynamical quenches. Similar studies have been performed in \cite{Rangamani:2015sha}, for a CFT driven by an oscillating composite scalar operator, and \cite{Biasi:2019eap} where a gapped holographic system has been perturbed with a homogeneous gravitational periodic driving. Some qualitative features observed in \cite{Rangamani:2015sha} are totally consistent with our findings.
\section{The Holographic Model}
We consider a $4$-dimensional holographic massive gravity model \cite{Baggioli:2014roa,Alberte:2015isw} defined by the following action:
\begin{equation}\label{action}
S\,=M_p^2\,\int d^4x \sqrt{-g}
\left[\frac{R}2+\frac{3}{\ell^2}- \, m^2 V(X)\right]
\end{equation}
with $X \equiv \frac12 \, g^{\mu\nu} \,\partial_\mu \phi^I \partial_\nu \phi^I$. The St\"uckelberg fields admit a radially constant profile $\phi^I\,=\,x^I$
which breaks the translational invariance of the dual field theory. For the rest of the manuscript, we focus on the specific potential $V(X)=X^3$, which realizes the spontaneous symmetry breaking of translations and it gives rise to a finite elastic response in the dual field theory and to the presence of propagating phonon modes -- the corresponding Goldstones \cite{Alberte:2017oqx,Ammon:2019apj,Baggioli:2019aqf,2019arXiv191005281B}.\footnote{See \cite{Baggioli:2019abx,Ammon:2019wci,Baggioli:2018vfc,Baggioli:2018nnp,Alberte:2017cch} for different choices of the potential $V(X)$ and the corresponding dual field theory properties.}  For the potential considered in this work, the sound speeds of transverse and longitudinal phonons are subluminal \cite{Alberte:2017oqx,Ammon:2019apj,Baggioli:2019elg}.  Moreover, the elastic response is accompanied by a viscous dissipative contribution \cite{Alberte:2016xja}, which qualifies the model as viscoelastic \cite{Andrade:2019zey}.\footnote{\color{black}In our model, the $ISO(2)$ global symmetry, typical of solids \cite{Nicolis:2015sra}, is not gauged in the bulk. It would be nice to understand better the physical consequences of this fact. For an elegant solution to this issue, see \cite{Esposito:2017qpj}.\color{black}} One direct consequence of the competition between elasticity and dissipation is the observation of a sound to diffusion crossover in the spectrum of transverse phonons \cite{Ammon:2019wci}, analogous to the Ioffe-Regel crossover in dissipative systems \cite{Baggioli:2018qwu}.\\
In the linear regime -- valid when the external deformations are small -- we can use linear response theory to obtain the shear correlator from the bulk theory using the holographic dictionary.
In the limit of zero momentum, the stress tensor correlator reads
\begin{equation}
    \mathcal{G}^R_{T_{xy}T_{xy}}(\omega,k=0)\,\equiv\,G'(\omega)\,+\,i\,G''(\omega) \label{linlin}
\end{equation}
and it defines for us the storage modulus $G'(\omega)$ and the loss modulus $G''(\omega)$, together with the loss angle (phase shift) $
    \tan \delta(\omega)\,\equiv\, \frac{G''(\omega)}{G'(\omega)}$. At low frequency we have:
    \begin{equation}
        \mathcal{G}^R_{T_{xy}T_{xy}}(\omega,k=0)\,=\,G_0\,-\,i\,\eta\,\omega\,+\,\mathcal{O}(\omega^2)
    \end{equation}
    where $G_0$ and $\eta$ are the static shear modulus and the shear viscosity, respectively.\\
In a perfect elastic solid, we have $G''=0$ and $\delta=0$, while in a purely dissipative fluid $G'=0$ and $\delta= \pi/2$. All the materials with $0<\delta<\pi/2$ are by definition viscoelastic.
\begin{figure}
    \centering
    \includegraphics[width=0.8 \linewidth]{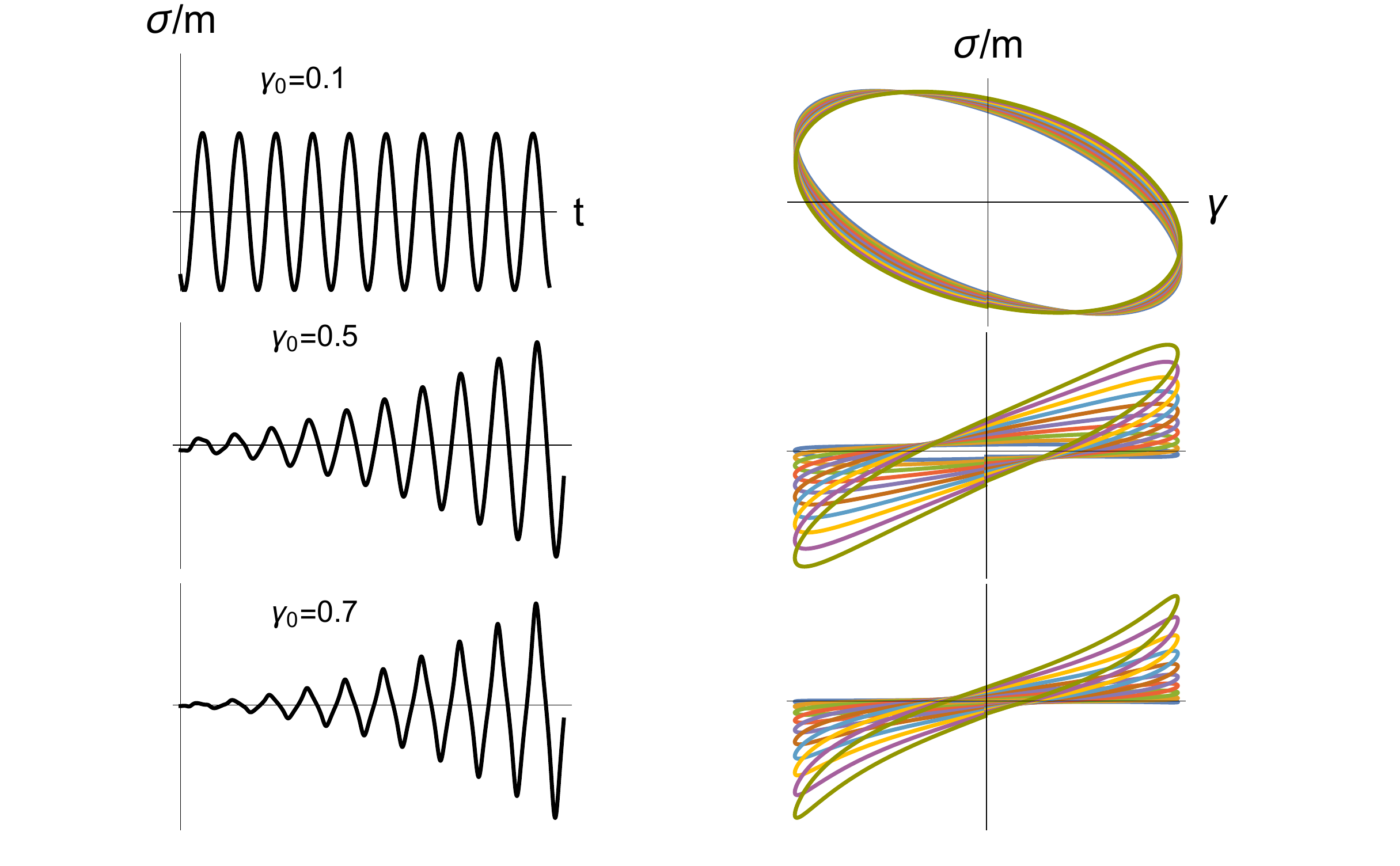}
    \caption{The onset of nonlinear elasticity by increasing the strain amplitude. The strain is $\gamma(t) = \gamma_0\,\sin (2\pi \omega t)$ with a smooth growing amplitude. Each colour
in the Lissajous figures correspond to the i$^{th}$ period. We
fix \color{black}$m/T_{in}=1.81$, $\omega/m = 0.32$.\color{black}}
    \label{fig:2}
\end{figure}\\
In the holographic model considered, at $m=0$, the static elastic modulus is null, $G_0=0$, and the system is a dissipative viscous fluid (saturating the Kovtun-Son-Starinets (KSS) bound, $\eta/s=1/4\pi$ \cite{Policastro:2001yc}).
At intermediate and finite $m/T$, the system has both a finite static modulus and a finite viscosity and it displays viscoelastic properties -- for details see the Supplementary Material (SM). The larger the parameter $m$ -- the mass of the graviton -- the stronger the elastic component.
\section{Nonlinear rheology}
\begin{figure}[t!]
    \centering
    \includegraphics[width=0.49 \linewidth]{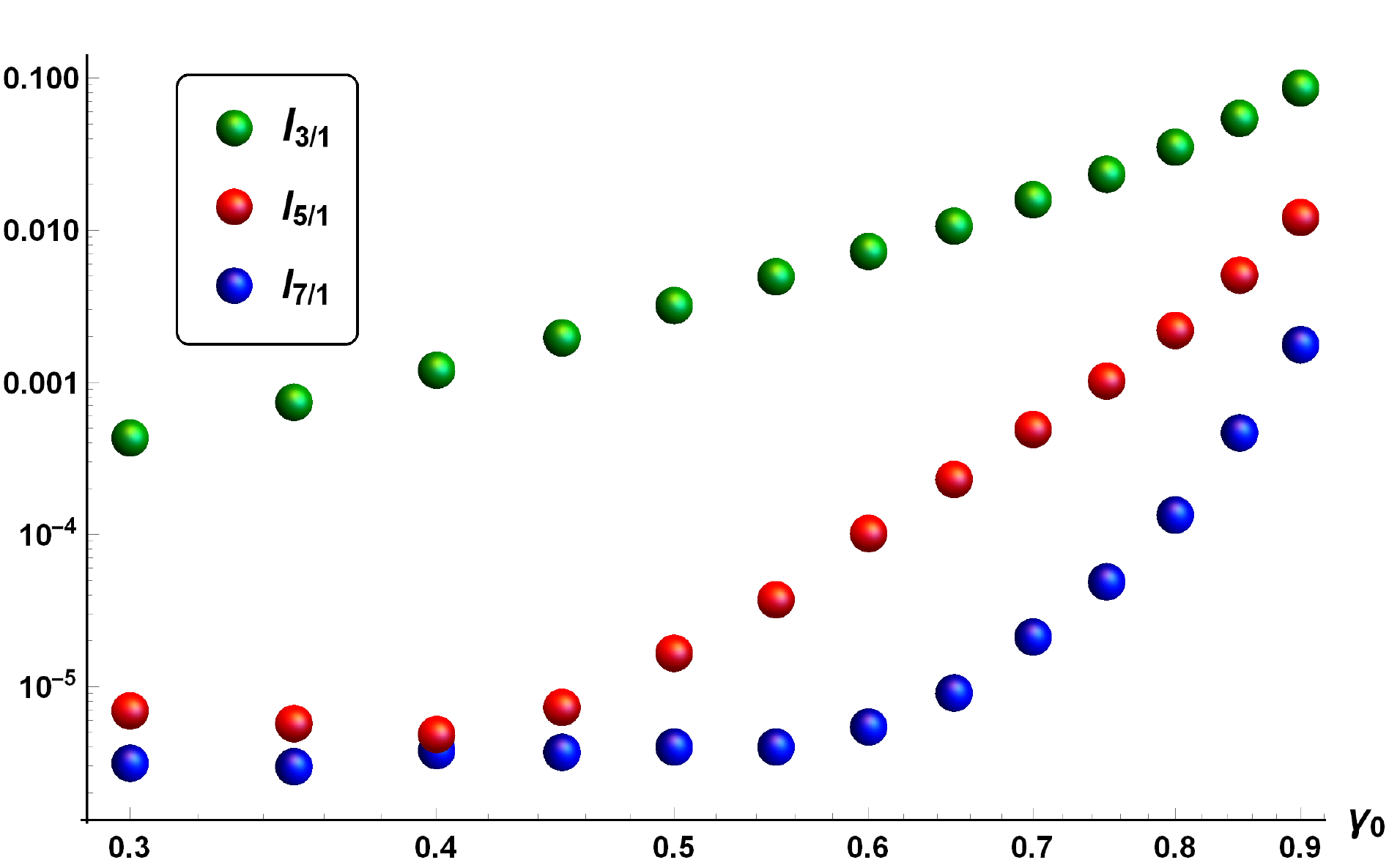}
    \includegraphics[width=0.49 \linewidth]{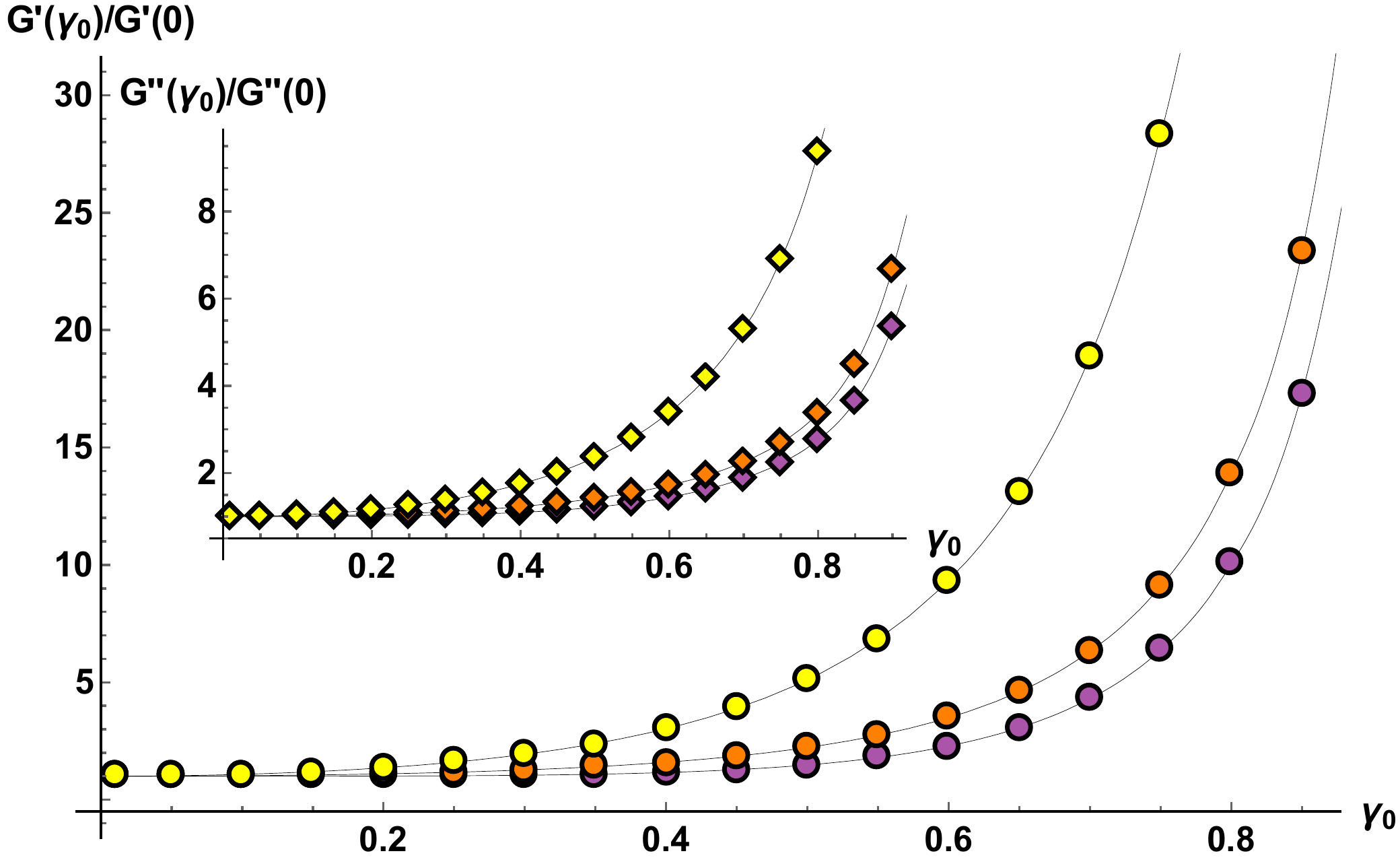}
    \caption{\textbf{Left: }The normalized intensity $I_{n/1}\equiv P(\omega_n)/P(\omega_1)$ of the first three higher harmonics in function of the strain amplitude $\gamma_0$. \textbf{Right: }The first complex moduli $G_1'(\omega,\gamma_0),G_1''(\omega,\gamma_0)$ at fixed frequency, in function of the strain amplitude. \color{black}We normalized them by their linear value $G(0)\equiv G(\gamma_0 \ll 1)$. \color{black} The colors correspond to $m/\color{black}T_{in}\color{black}=0.01,1.81,30$ (from yellow to purple). The onset of nonlinearity is roughly independent of the value of $m/\color{black}T_{in}\color{black}$ and it appears around $\gamma_0 \sim 0.3$.}
    \label{figbb}
\end{figure}

Whenever the amplitude of the applied strain is large, nonlinearities set in and the linear viscoelastic approximation fails. From a gravitational point of view, this problem requires a more complicated time-dependent setup which is explained in detail in  the SM , following the seminal work of \cite{Chesler:2013lia}.\footnote{\color{black} Importantly, out-of-equilibrium,  temperature is not a well defined concept \cite{2017PhR...709....1P}. For this reason, we will indicate it everywhere with the symbol $T_{in}$ to specify that such parameter is the temperature of the initial ($t=0$) state.\color{black}} Within this regime, the produced stress is no longer linearly proportional to the applied strain but it presents a distorted shape which can be understood as a superposition of different Fourier components. More specifically, in the nonlinear regime, the strain $\gamma$ and the stress $\sigma$ can be represented as\footnote{The reason why only odd powers appear in the expansion is that the stress response is typically taken independent of
the shear direction.}
\begin{align}
    &\gamma(t)\,=\,\gamma_0\,\sin(2 \pi \omega t)\,,\quad \dot{\gamma}(t)\,=2\pi\omega\, \,\gamma_0\,\cos(2\pi \omega t)\\ &\sigma(t)\,=\,\sum_{p,odd} \sum_{q,odd}^p \gamma_0^q\,\left(a_{pq}\sin (2\pi q \omega t)+ b_{pq}\,\cos(2\pi q \omega t)\right)\label{dd}
\end{align}
where $a_{11},b_{11}$ correspond to the complex moduli $G'(\omega),G''(\omega)$ in the linear regime, and the first nonlinear corrections entering at order $\mathcal{O}(\gamma_0^3)$.\\
In this letter, we will explore different methods to represent and characterize the nonlinear response at large amplitudes: (I) the analysis of the Fourier spectrum of the time dependent stress response, (II) the Lissajous figures -- stress-strain parametric curves $\{\gamma(t),\sigma(t)\}$, (III) the definition of the nonlinear complex moduli and their dependence on the strain amplitude.\\ 
\color{black}In the rest of the manuscript all quantities are displayed in units of the mass parameter $m$. Moreover, we fixed $h_{xx}=h_{yy}=1$ (see the SM and eq.(A.21) for details). Given this assumption, the strength of the applied external strain lies in the range $\gamma_0 \in [0,1]$.\color{black} 
\begin{figure}
    \centering
    \includegraphics[width=0.7 \linewidth]{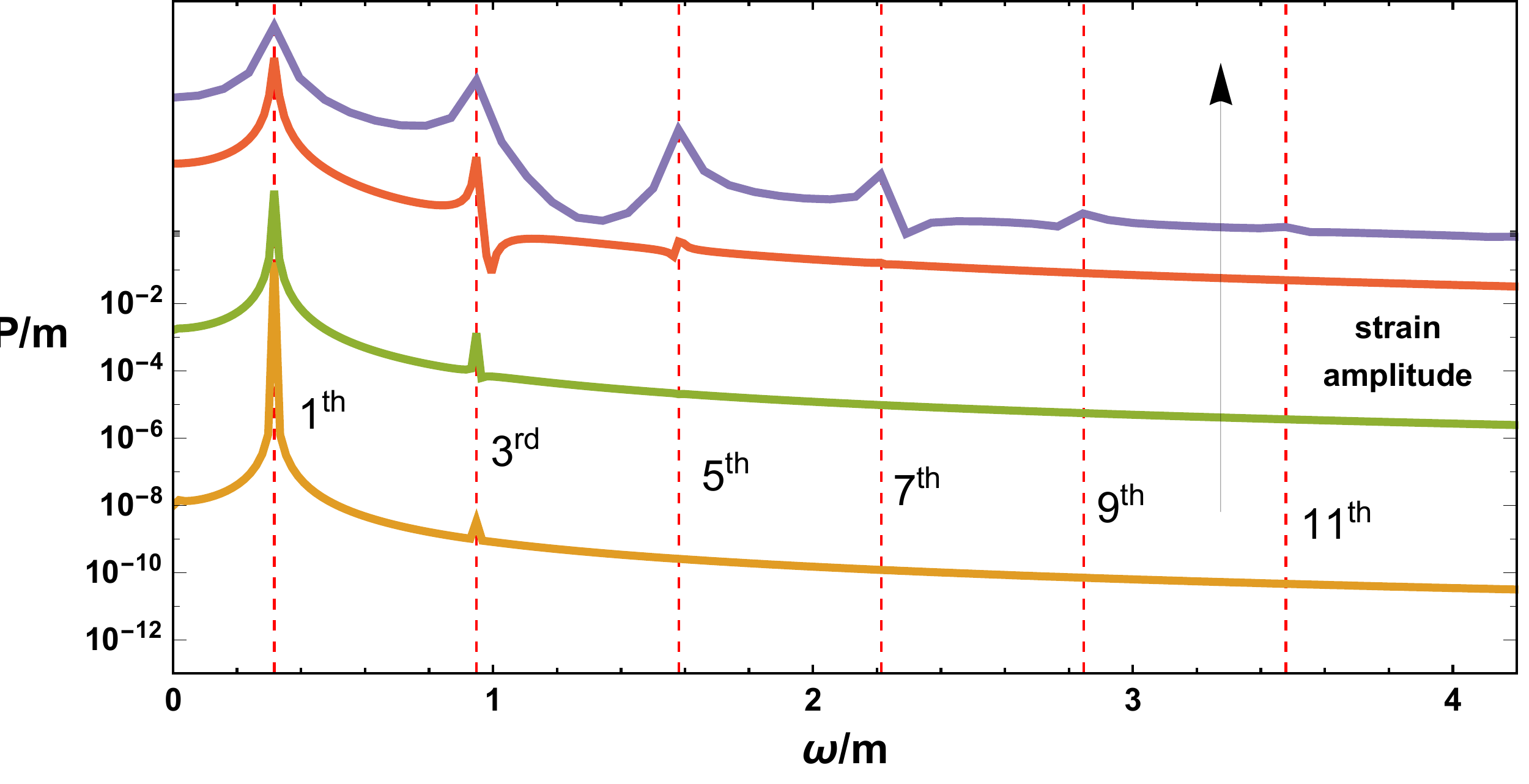}
    \caption{Fourier spectrum $\mathcal{P}$ of the time dependent stress for increasing strain amplitude $\gamma_0=\{0.01,0.1,0.4,0.75\}$ (from orange to blue). Increasing the strain amplitude higher (odd) harmonics appear.  The power spectrum is defined as $\mathcal{P}(\omega)\equiv \mathcal{F}[\,\int_{-\infty}^{\infty}\sigma(t+\tau)\sigma(t)dt\,]$.}
    \label{fig:3b}
\end{figure}\\
Firstly, we observe in fig.\ref{fig:2} that by increasing the amplitude of the applied strain the shape of the stress response gets distorted and it deviates from a simple oscillatory function. This behavior is also displayed in the corresponding Lissajous figures which are no longer a simple oval, as expected in the linear regime. We notice that the shape of the curve after each cycle appears to be slightly modified; this phenomenon emphasizes the complexity of our viscoelastic system.\\
In fig.\ref{fig:3b} we study the Fourier spectrum of the signal. At small amplitude (orange curve), the spectrum is localized on the first and only harmonic, which is fixed by the frequency of the applied strain signal. This means the system is still in the linear response regime, where the stress is linearly proportional to the applied strain. By increasing the amplitude, higher (odd) harmonics appear in the spectrum confirming the functional structure displayed in eq.\eqref{dd}. The normalized power of the higher harmonics, $I_{n/1}$ is shown in fig.\ref{figbb}.  We find preliminary evidence for a  power law behavior $\sim\gamma_0^{n}$ which was previously suggested by theoretical arguments in \cite{doi:10.1146/annurev.fl.27.010195.001125}.
\\
Continuing along the lines of eq.\eqref{dd}, we can rewrite the stress response as:
\begin{equation}
    \sigma(t)=\sum_{n,odd} \sum_{m,odd}^n \gamma_0^n \left(G'_{nm}\sin (2\pi m \omega t)+G''_{nm}\cos (2\pi m \omega t)\right)\label{ee}
\end{equation}
The complex moduli are rigorously defined only in the linear regime; however, the measurements of $G'(\gamma_0)$ and $G''(\gamma_0)$ at a fixed frequency can provide meaningful information. The most common option to calculate the moduli from a 
non-sinusoidal response consists in looking at the quantities $G_1'(\omega,\gamma_0),G_1''(\omega,\gamma_0)$, defined as the contributions from the first harmonics $\sin (2\pi\omega t), \cos (2\pi\omega t)$ 
 to the expansion in eq.\eqref{ee}. Additionally, the values of $G'_1,G''_1$ are exactly what the commercial rheometers provide in the experiments. Using a simple expansion, we obtain the first term in the sum of eq.\eqref{ee}:
\begin{align}
    \sigma(t)^{\textit{first term}}=&\left[G'_{11}\,\gamma_0+G'_{31}\,\gamma_0^3+\mathcal{O}(\gamma_0^5)\right]\sin (2\pi \omega t)\nonumber\\
    +&\left[G''_{11}\,\gamma_0+G''_{31}\,\gamma_0^3+\mathcal{O}(\gamma_0^5)\right]\cos (2\pi \omega t)\nonumber\\
    =\,G'_1(\omega,\gamma_0)&\gamma_0 \sin (2\pi \omega t)+G''_1(\omega,\gamma_0)\gamma_0\cos (2\pi \omega t)
\end{align}
where we are neglecting the higher harmonics corrections which naturally appear in eq.\eqref{ee}. Given these notations, the values of $G_1',G_1''$ at zero strain correspond to the linear response limit in eq.\eqref{linlin}.
We plot the dependence of the first nonlinear complex moduli $G_1'(\omega,\gamma_0), G_1''(\omega,\gamma_0)$ at fixed frequency in fig.\ref{figbb}. We observe that for small amplitudes the moduli are independent of the strain amplitude. This is not true anymore at large amplitudes where nonlinear effects become important. 
 We find that the onset of nonlinearity is roughly independent of the value of $m/\color{black}T_{in}\color{black}$ and it depends solely on the amplitude of the applied strain $\gamma_0$. Notice that in the nonlinear regime both moduli grow in a faster-than-linear fashion. From an operational point of view, this defines the presence of \textit{strain stiffening}. This behavior is typical of hyperelastic materials such as rubber-like systems or complex polymers and it is in contrast to the so-called \textit{strain hardening} which is on the contrary a common feature of rigid metals.\\
Let us also notice that at low strain, for small values of the mass $m$, $G_1''>G_1'$, indicating that our dual field theory is a \textit{viscoelastic liquid}. This is reversed at large values of $m/\color{black}T_{in}\color{black}$, where the system becomes a \textit{viscoelastic solid} with $G_1''<G_1'$  \cite{PhysRevLett.118.018002} (see fig.(A.2) in the SM). This is totally consistent with the fact that the graviton mass $m$ determines the ''amount of solidity'' of the system -- its rigidity.\\
A second possibility to characterize the nonlinear response, which is explored in detail in the SM, consists in defining the complex moduli from the Lissajous figures looking at the tangent and the secant of the curve. Using this second method, we consistently find that the small amplitude modulus is smaller than the large amplitude one (see fig.(A.6) in the SM ) confirming the \textit{strain stiffening} scenario.
\begin{figure}[t!]
    \centering
    \includegraphics[width=0.8 \linewidth]{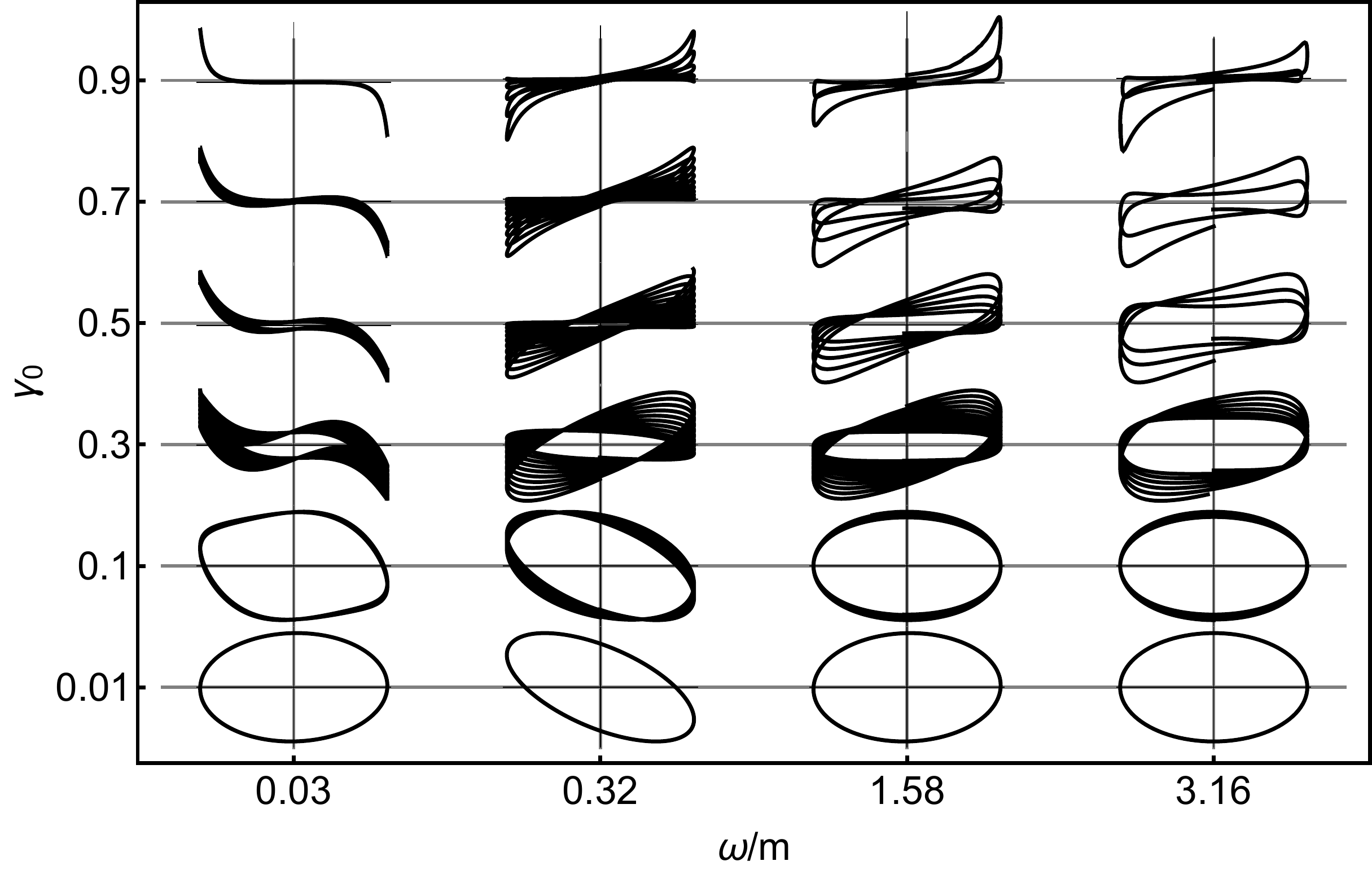}
    \caption{Pipkin diagram: the Lissajous figures in function of the amplitude and the frequency of the oscillatory strain source. For this plot we fix $m/\color{black}T_{in}\color{black}=1.81$. This choice relates to a regime where our system is concretely viscoelastic.}
    \label{fig:3w}
\end{figure}\\
To complete our analysis, we construct the Pipkin diagram of our model in fig.\ref{fig:3w} by plotting the Lissajous figures at various strain frequencies and amplitudes. We observe a neat transition between a linear viscoelastic regime at low amplitude and frequency to a more complicated large regions where the response becomes highly nonlinear (notice the similarities with \cite{Rangamani:2015sha}). This last result confirms that the regime we investigated cannot be described by linear response and it displays all the main physical properties of LAOS systems.
\section{Discussion}
In this letter, we characterize the nonlinear and time dependent mechanical response of viscoelastic (strongly coupled) field theories using holographic techniques. We focus our analysis on oscillatory external strains and on the regime of LAOS. We prove that the viscoelastic response is quite similar to that of complex fluids and hyperelastic materials (e.g. rubber) and in particular it exhibits a very neat \textit{strain stiffening} phenomenon (see \cite{Erk2010} for some concrete experimental data), which suggests that the holographic models at hand are not suitable to describe ordered metallic crystals characterized by \textit{strain hardening}. We also observe a transition between a \textit{viscoelastic liquid} behavior at small graviton mass, $m/T \ll 1$, to a \textit{viscoelastic solid} regime at large values, which confirms the identification of the graviton mass with the rigidity of the dual field theory.\\
This work opens a new path for the study of complex fluids and viscoelastic systems using the holographic methods, which so far have been successfully applied only to strongly coupled liquids with no elastic response.\\
There are several direct and interesting directions to pursue. Firstly, it would be desirable to reach a better theoretical understanding of our numerical data by
comparing our results to known phenomenological models such as the multi-mode Giesekus model \cite{doi:10.1146/annurev.fl.27.010195.001125,doi:10.1063/1.2964623}.\\
Secondly, a more extensive exploration of the phase diagram and possibly an extension of the study to include
also the Chebyshev analysis \cite{cho2018viscoelasticity} are certainly needed to draw universal conclusions.\\
On a more phenomenological perspective, one could consider different types of experiments, \textit{i.e.} different signals for the applied strain such as building up functions, step functions and quenches. This extension would permit the study of extremely interesting phenomena such as nonlinear relaxation, stress overshoot, yielding, which represent still open challenges for rheology and condensed matter in general.  In this respect, the existence of possible universal (relaxation) timescales is certainly a fundamental question to answer.\\
Finally, another relevant question which our work poses is the possibility of having holographic models displaying \textit{strain hardening} and being therefore more suitable to describe metallic solids. This nicely connects to a fundamental and still open question: which kind of solids do these holographic systems describe? As shown in this letter, the analysis of nonlinear transport properties is certainly a way of resolving this conundrum.
\section*{Acknowledgements}
We thank Martin Ammon, Alex Buchel, Victor C.Castillo, Romulad Janik, Andreas Karch, Marco Laurati, Javier Mas, Ayan Mukhopadhyay, Oriol Pujolas, Yuho Sakatani, Laurence Yaffe and  Alessio Zaccone for interesting discussions and useful comments about the topic of the manuscript and an early version of this draft. S.G. thanks Martin Ammon for the collaboration on a related project. M.B.acknowledges the support of the Spanish MINECO’s “Centro de Excelencia Severo Ochoa” Programme under grant SEV-2012-0249. S.G. gratefully acknowledges financial support by the DAAD (German Academic Exchange Service) for a \textit{Jahresstipendium f\"ur Doktorandinnen und Doktoranden} (One-Year Research grant for doctoral candidates) in 2019.

\bibliographystyle{apsrev4-1}
\bibliography{nonlinear}
\onecolumngrid

\clearpage
\newpage
\twocolumngrid
\appendix
\renewcommand\thefigure{A.\arabic{figure}}    
\setcounter{figure}{0} 
\renewcommand{\theequation}{A.\arabic{equation}}
\setcounter{equation}{0}
\section*{Supplementary material}
\subsection{Linear response}\label{app1}
We review the linear viscoelastic analysis of the holographic model used in the main text and defined from the bulk action in eq.\eqref{action}. More details about this linear regime can be found in \cite{Andrade:2019zey}.\\
Given an applied external oscillatory strain $\gamma(t)=\gamma_0 \sin (\omega t)$, the linear viscoelastic response is encoded in a time dependent shear stress of the form:
\begin{equation}
    \sigma(t)\,=\,\gamma_0\,\left(G'(\omega)\,\sin (\omega t)\,+\,G''(\omega)\,\cos(\omega t)\right)
\end{equation}
where $G'(\omega),G''(\omega)$ are the storage and loss moduli, determining the elastic in-phase response and the dissipative out-of-phase one.\\
Using the Kubo formulae formalism, these two moduli can be read from the shear stress tensor two-point functions as follows:
\begin{equation}
     \mathcal{G}^R_{T_{xy}T_{xy}}(\omega,k=0)\,\equiv\,G'(\omega)\,+\,i\,G''(\omega)
\end{equation}
The shear correlator can be derived with standard techniques using the holographic dictionary and considering the bulk dynamics of a gravitational wave perturbation -- a geometric perturbation $\delta g_{xy}$.
\begin{figure}[h!]
    \centering
    \includegraphics[width=0.7 \linewidth]{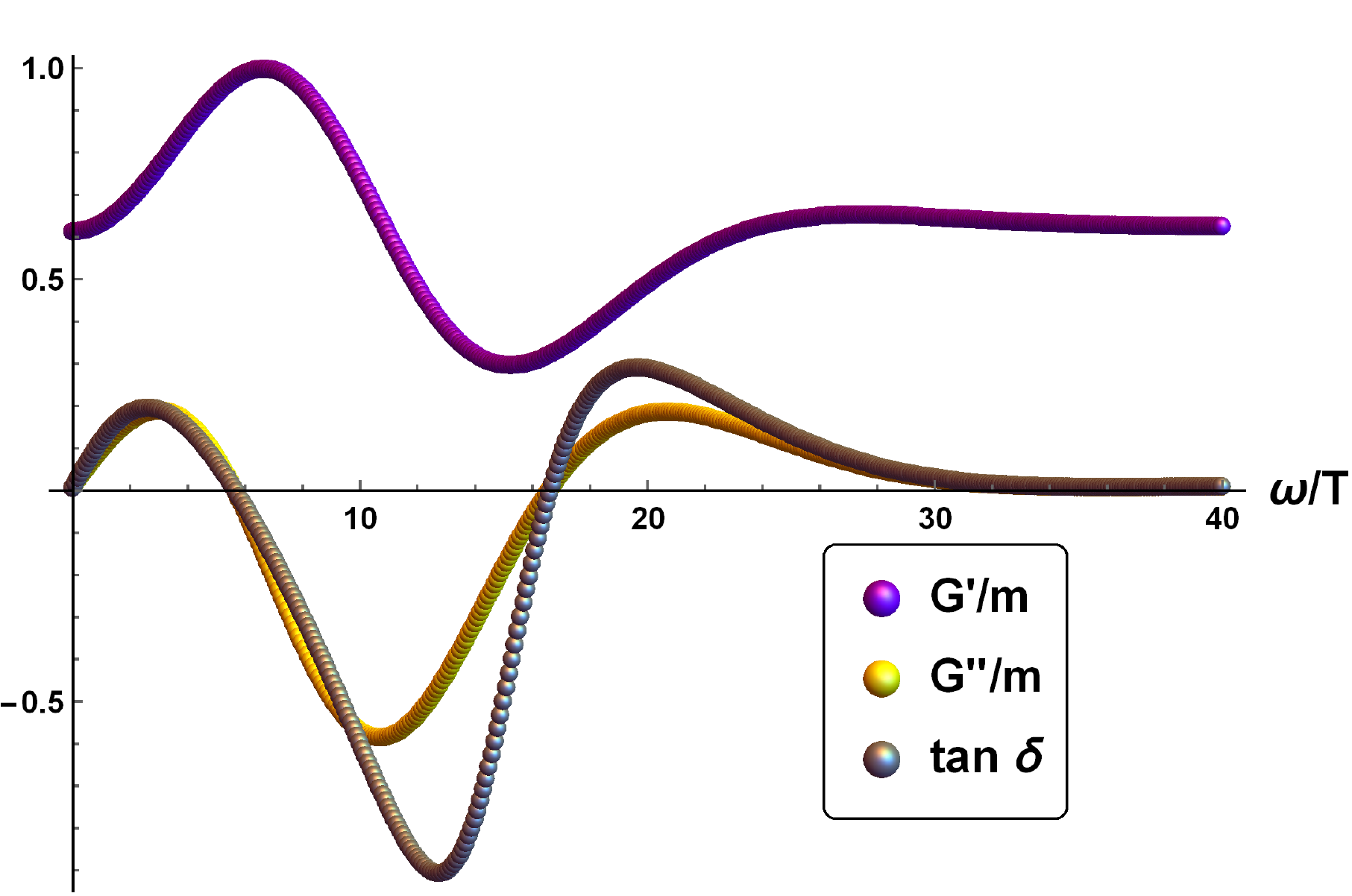}
    
    \vspace{0.4cm}
    
    \includegraphics[width=0.7\linewidth]{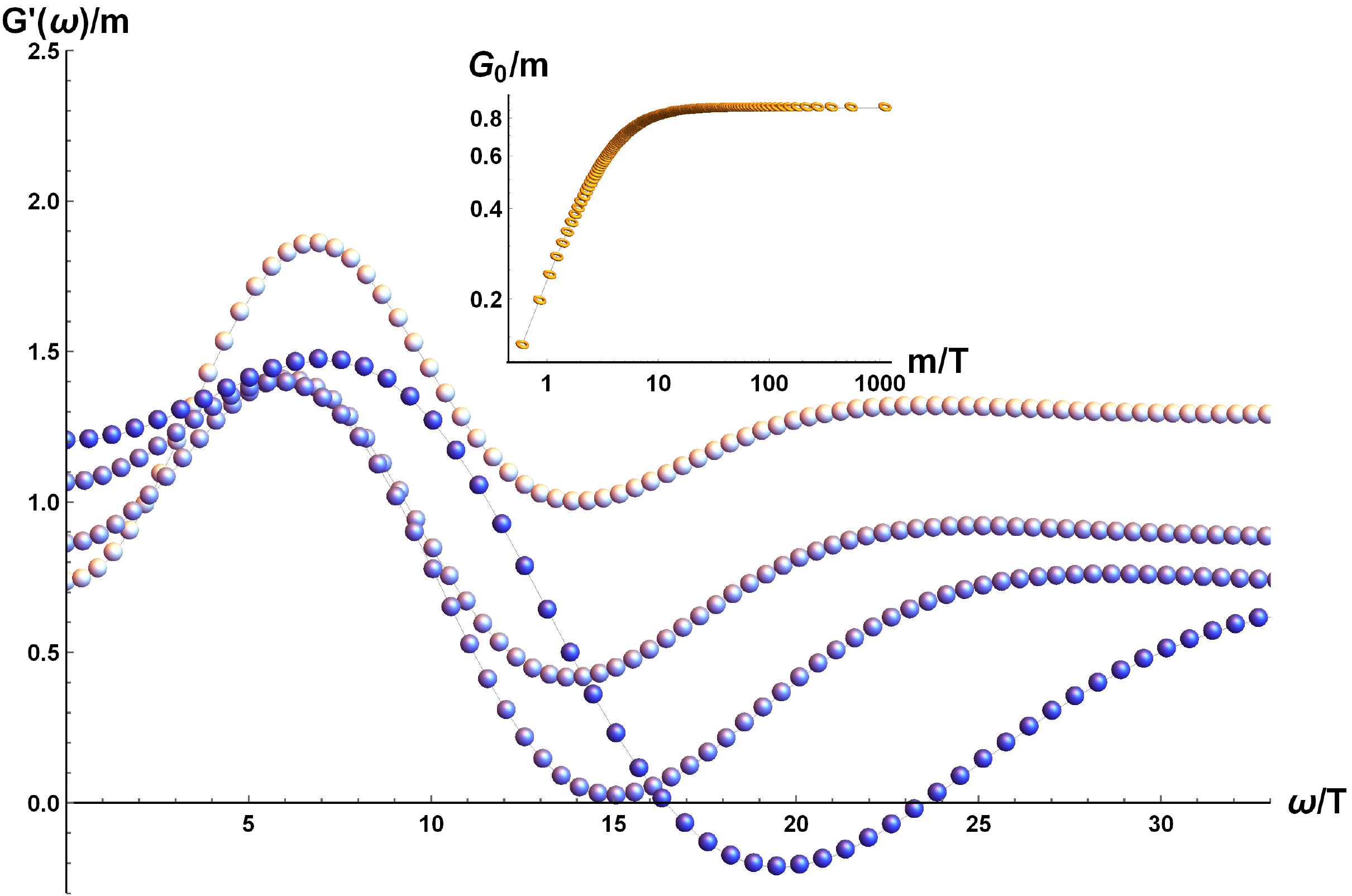}
    \caption{\textbf{Top: }The linear viscoelastic regime. The storage and loss moduli $G'(\omega),G''(\omega)$ and the loss angle $\delta\equiv G''/G'$ in function of the frequency $\omega$. We fix $m/T=3.55$. At low frequency, we have $G'(\omega)=G_0,G'(\omega)=\omega \eta$; at large frequency $G'(\omega)=G_\infty,\,G'(\omega)=0$. \textbf{Bottom: }The storage modulus $G'(\omega)$ for increasing values of $m/T$ (from light to dark blue). The inset shows the value of the static shear modulus $G_0\equiv G'(0)$ in function of $m/T$. The limiting value $m/T=0$ corresponds to the Schwarzschild solution.}
    \label{fig:0b}
\end{figure}\\
Concretely, we consider a four dimensional asymptotically AdS black hole geometry in Eddington-Filkenstein (EF) coordinates:
\begin{equation}
\label{backg}
ds^2=\frac{1}{u^2} \left[-f(u)\,dt^2-2\,dt\,du + dx^2+dy^2\right]
\end{equation}
where $u\in [0,u_h]$ is the radial holographic direction spanning from the boundary $u=0$ to the horizon $u=u_h$. The emblackening factor appearing in the BH geometry takes the simple form:
\begin{equation}\label{backf}
f(u)= u^3 \int_u^{u_h} dv\;\left[ \frac{3}{v^4} -\frac{m^2}{v^4}\, 
V(v^2) \right] \, ,
\end{equation}
and the corresponding temperature of the dual field theory reads:
\begin{equation}
T=-\frac{f'(u_h)}{4\pi}=\frac{6 -  2 m^2 V\left(u_h^2,u_h^4 \right) }{8 \pi u_h}~.
\end{equation}
Additionally, the entropy density is given by $s=2\pi/u_h^2$ and the heat capacity can be directly obtained as $c_v=T ds/dT$.
As already explained, the linear viscoelastic properties of the model are encoded in the dynamics of the $T_{xy}$ operator which is holographically dual to the bulk perturbation $\delta g_{xy}\equiv h_{xy}$.
At zero momentum $k=0$, the linearized equation for the field $h_{xy}$ decouples and in EF coordinates reads:
\begin{equation}
h_{xy}\left(-\frac{2 m^2 V_X}{f}-\frac{2 i \omega}{u f}\right)+h_{xy}' \left(\frac{f'}{f}+\frac{2 i \omega}{f}-\frac{2}{u}\right)+h_{xy}''=0\label{EQ}
\end{equation}
where $V_X\equiv \partial_X V(X)$ and primes denote radial derivatives. 
The UV asymptotic behavior of the $h_{xy}$ field is:
\begin{equation}
h_{xy}\,=h_{xy\,(l)}(\omega)\,(1\,+\,\dots)\,+\,h_{xy\,(s)}(\omega)\,u^{3}\,(1\,+\,\dots)
\end{equation}
Finally, we can read the shear Green function from
\begin{equation}
 \mathcal{G}^{\textrm{(R)}}_{T_{xy}T_{xy}}(\omega)\,=\,\frac{2\,\Delta-d}{2}\,\frac{h_{xy\,(s)}(\omega)}{h_{xy\,(l)}(\omega)}\,=\,\frac{3}{2}\frac{h_{xy\,(s)} (\omega) }{h_{xy\,(l)} (\omega)}\label{green}
\end{equation}
which is the fundamental relation to characterize the linear viscoelastic response.
\begin{figure}[h!]
    \centering
    \includegraphics[width=0.7 \linewidth]{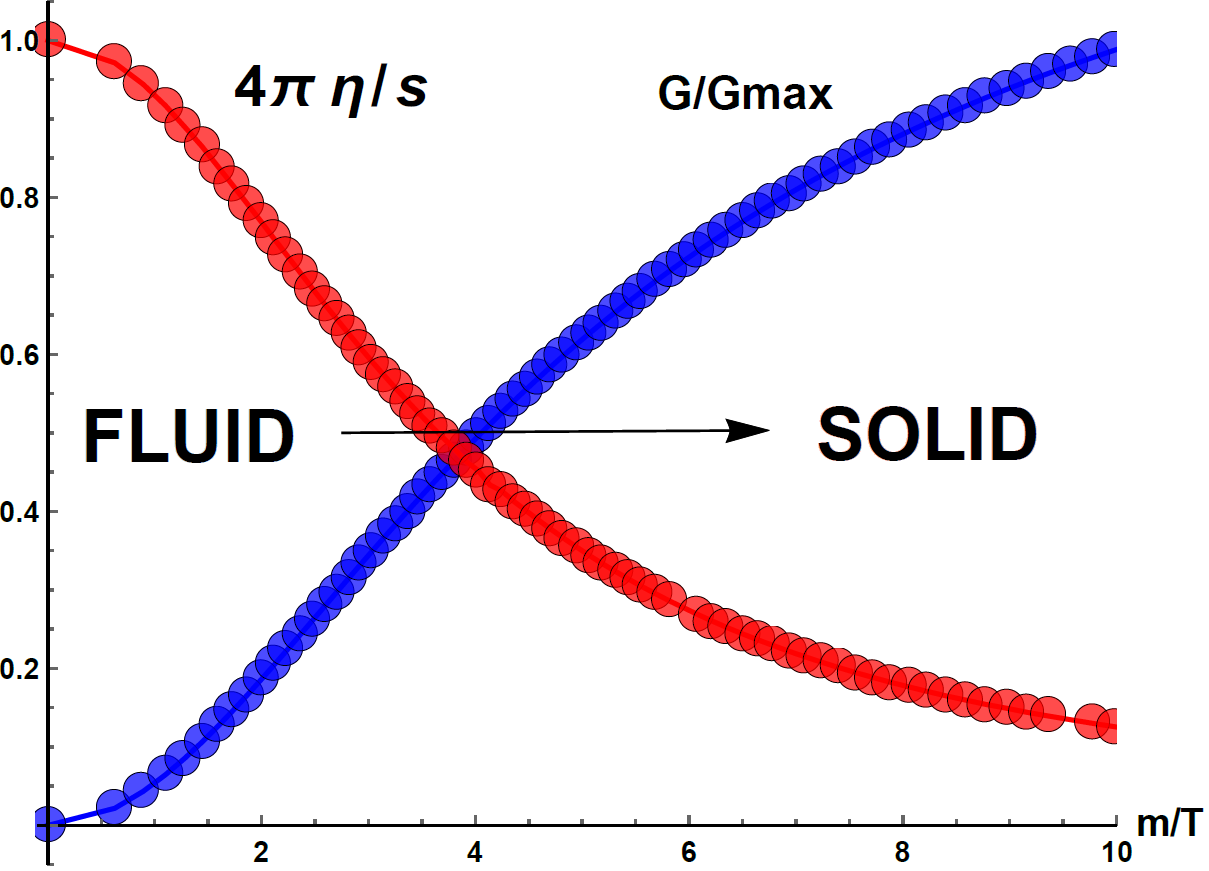}
    \caption{The normalized viscosity to entropy ratio and the normalized static elastic modulus ($G_{max}\equiv G_0(T=0)$) in function of the dimensionless parameter $m/T$. Figure taken and adapted from \cite{Andrade:2019zey}.}
    \label{fig:trans}
\end{figure}\\
In the low frequency region:
\begin{equation}
    G'(\omega)\,=\,G_0\,,\quad G''(\omega)\,=\,\eta\,\omega
\end{equation}
where $G_0$ is the static shear modulus and $\eta$ the shear viscosity; in the opposite limit of large frequency:
\begin{equation}
    G'(\omega)\,=\,G_{\infty}\,,\quad G''(\omega)\,\rightarrow\,0
\end{equation}
where $G_\infty$ is the instantaneous shear modulus.\\
At small values of $m/T$, the static shear modulus can be computed analytically \cite{Alberte:2016xja,Alberte:2017oqx}, using formula:
\begin{equation}
    G_0\,=\,m^2\,\int_0^{u_h}\,\frac{V'(\zeta^2)}{\zeta^2}\,d\zeta\,+\,\mathcal{O}(m^4)
\end{equation}
At the same time, the instantaneous shear modulus is universally given by \cite{Andrade:2019zey} :
\begin{equation}
    G_{\infty}\,=\,\frac{3}{8}\,\varepsilon
\end{equation}
where $\varepsilon$ is the energy density of the system.\\
The only dimensionless and relevant physical parameter of our setup is the ratio $m/T$; the physics of the system depends solely on its value. At small values of $m/T$, the viscosity of the system is larger than its elasticity -- we are in a \textit{viscoelastic liquid} regime. At an intermediate value of $m/T\sim \mathcal{O}(1)$, the situation is inverted. The system becomes more and more rigid and it enters the \textit{viscoelastic solid} regime.\\
\color{black}Notice that, in this section, since we are only considering the linear response regime, the parameter $T$ is the well defined temperature of the dual field theory.\color{black}
\subsection{Details about the nonlinear computation and the holographic renormalization}\label{app2}
The simplest time dependent metric in Eddington-Finkelstein parametrization with $g_{xy}\neq0$ and homogeneous symmetry is
\begin{eqnarray}
&&ds^2=-A(u,t)dt^2-\frac{2 du dt}{u^2}+S(u,t)^2\\
&&\left( \cosh{\left(H(u,t)\right)} (dx^2+dy^2)+2\sinh{\left(H(u,t)\right)} dx dy \right)\nonumber
\end{eqnarray}
To implement the characteristic formulation \cite{Chesler:2013lia}, we use hyperbolic functions such that $\text{Volume}_{xy}=S^2$. This choice leads to the following equations of motion,
\begin{eqnarray}
&&S{}^{''} +\frac{2}{u}S{}^{'}+\frac{{H{}^{'}}^2 }{4}S=0,\label{constraint}\\
&&{d_+S}{}^{'} +\frac{S{}^{'} }{S } {d_+S}=-\frac{3S}{2 u^2} +4m^2   \frac{\cosh^3 (H)}{u^2S^5},\label{eq-dpS}\\
&&{d_+H}{}^{'} +\frac{ S^{'} }{S}d_+H=-\frac{H{}^{'} {d_+S} }{S }\label{eq-dpH}\\
&&\hspace{80pt}-24m^2 \frac{\sinh (H) {\cosh^2 (H)}}{u^2 {S^{6}}},\nonumber\\
&&A{}^{''}+\frac{2 }{u}A{}^{'}= \frac{{d_+H}  H{}^{'} }{u^2 }-\frac{4S{}^{'}  {d_+S}  }{u^2 S^2}\label{eq-A}\\
&&\hspace{40pt}+48m^2 \frac{  \cosh^3 (H)}{u^4S^{6}},\nonumber\\
&&4{d_+^2S} +2 u^2 A{}^{'}d_+S+ {d_+H}^2 S =0,\label{eq-dppS}
\end{eqnarray}
where $d_+\mathcal{A}:= \dot{\mathcal{A}}-\frac{A}{u^2}\mathcal{A}^{'}$ is the directional derivative, and prime stands for derivative with respect to radial coordinate $u$ and we use dot for derivative with respect to time $t$. 

The metric functions have the following near boundary expansion\footnote{Using the residual symmetry in EF radial coordinate, we fix the apparent horizon at $u_h=1$ and leave $s_1$ as a dynamical parameter.}
\begin{eqnarray}
&&A=\frac{1}{u^2}+\frac{2(s_1-\dot{s_0})}{s_0 \ u}+\left( \frac{s_1^2}{s_0^2}-\frac{2\dot{s_1}}{s_0}-\frac{3\dot{h_0}^2}{4} \right)+a_3 u+\mathcal{O}(u^2),\nonumber\\
&&S=\frac{s_0}{u}+s_1-\frac{s_0\dot{h_0}^2}{8}u+\frac{s_1 \dot{h_0}^2}{8} u^2+\mathcal{O}(u^3),\label{bdry-exp}\\
&&H=h_0+\dot{h_0}u-\frac{s_1 \dot{h_0}}{s_0}u^2+h_3 u^3+\mathcal{O}(z^4).\nonumber
\end{eqnarray}
accompanied with the Ward identity,
\begin{eqnarray}
 &&\dot{a_3}+\frac{3 a_3 \dot{s_0}}{s_0}-\dot{h_0}{}^2 \left(\frac{\dot{s_0}{}^2-3 s_1{}^2}{2 s_0{}^2}+\frac{\ddot{s_0}}{2 s_0}\right)-\frac{3 \dot{h_0} \ddot{h_0} \dot{s_0}}{2 s_0}\nonumber\\
 &&+\frac{3}{8} \dot{h_0}{}^4-\frac{3}{2} h_3 \dot{h_0}-\frac{1}{2} \dddot{h_0} \dot{h_0}=0.
\end{eqnarray}

Since we want to impose the shear strain as the source for the boundary metric $h_{xy}=\gamma(t)$  and  keep the spatial length scale of the boundary theory fixed $h_{xx}=h_{yy}=1$, we use 
\begin{equation}
    h_0(t)=\!\text{arcsinh}\left(\frac{\gamma(t)}{\left(1-\gamma(t)^2\right)^{1/2}}\right),\  s_0(t)=\left(1-\gamma(t)^2\right)^{1/4}.
\end{equation}
Note that the strain function has to be less than one, in our units, to have real metric functions.

Using the standard holographic renormalization, one can read the boundary  stress tensor  from the near boundary expansion of the metric components as\footnote{From near boundary expansion \eqref{bdry-exp} it's clear that the back reaction of the scalar fields in the metric components does not contribute in the stress tensor.}
\begin{eqnarray}\label{stress-tensor}
&&T_{tt} = -a_3,\\
&&T_{xx} = -\frac{{a_3}}{2} - \frac{3  \gamma  {s_1}^2 \dot{\gamma}}{2 \left(1- \gamma ^2\right)^{3/2}}+\frac{3 h_3}{2}\\
&&-\frac{ \gamma  \left(4 \left( \gamma ^2-1\right)^2 \dddot{\gamma}+3 \left(4  \gamma ^2+1\right) \dot{\gamma}^3-16  \gamma  \left( \gamma ^2-1\right) \dot{\gamma} \ddot{\gamma}\right)}{8 \left( \gamma ^2-1\right)^3}  \gamma,\nonumber \\
&&T_{xy} = -\frac{a_3}{2}  \gamma - \frac{3 {s_1}^2 \dot{\gamma}}{2 \left(1- \gamma ^2\right)^{3/2}} + \frac{3 {h_3}}{2} \\
&&- \frac{4 \left( \gamma ^2-1\right)^2 \dddot{\gamma}+3 \left(4  \gamma ^2+1\right) \dot{\gamma}^3-16  \gamma  \left( \gamma ^2-1\right) \dot{\gamma} \ddot{\gamma}}{8 \left( \gamma ^2-1\right)^3}.\nonumber
\end{eqnarray}
The shear stress to the strain function $h_{xy}(t)=\gamma(t)$ is given by off-diagonal component of the energy-momentum tensor $T_{xy}(t)$.

\subsection{Numerical routine}\label{app3}
To impose the boundary conditions, it is convenient to use the following redefined tilde functions
\begin{eqnarray}
&&A=\frac{1+\tilde{A}\ u}{u^2},\quad S=\frac{s_0+\tilde{S}\ u}{u},\nonumber \\
&&H=h_0+\tilde{H}\ u,\quad d_+S=\frac{\widetilde{d_+S}}{u^2},\\
&&d_+H=\frac{\dot{h_0}}{2}+\widetilde{d_+H}\ u.\nonumber
\end{eqnarray}
In our numerical calculation we use the Chebyshev discretization with 50-100 grid points for integration  along the radial coordinate.
\begin{figure}[h!]
    \centering
    \includegraphics[width=0.6 \linewidth]{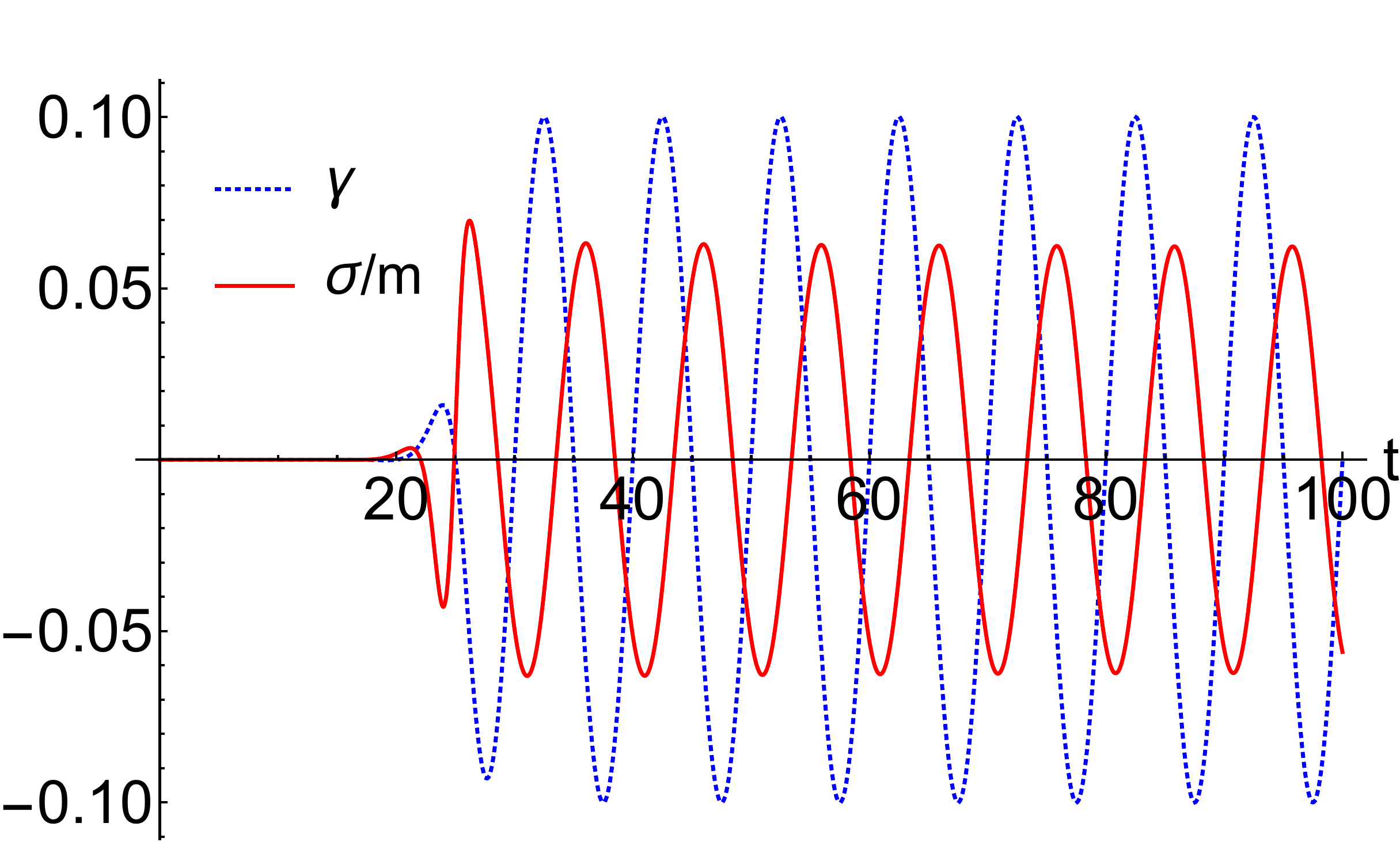}
    
    \vspace{0.15cm}
    
    \hspace{-0.06 \linewidth}\includegraphics[width=0.47 \linewidth]{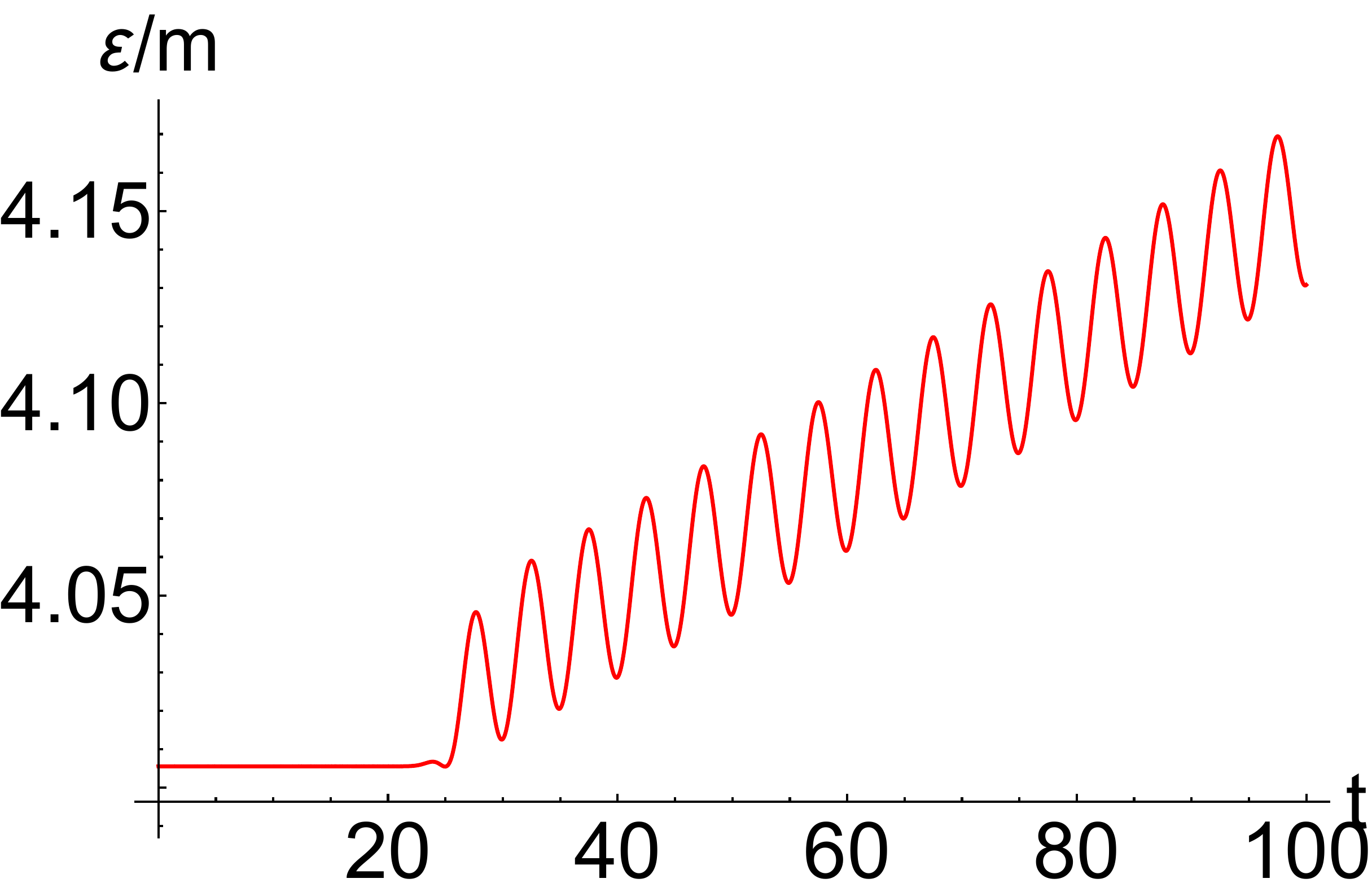}
\quad \includegraphics[width=0.47\linewidth]{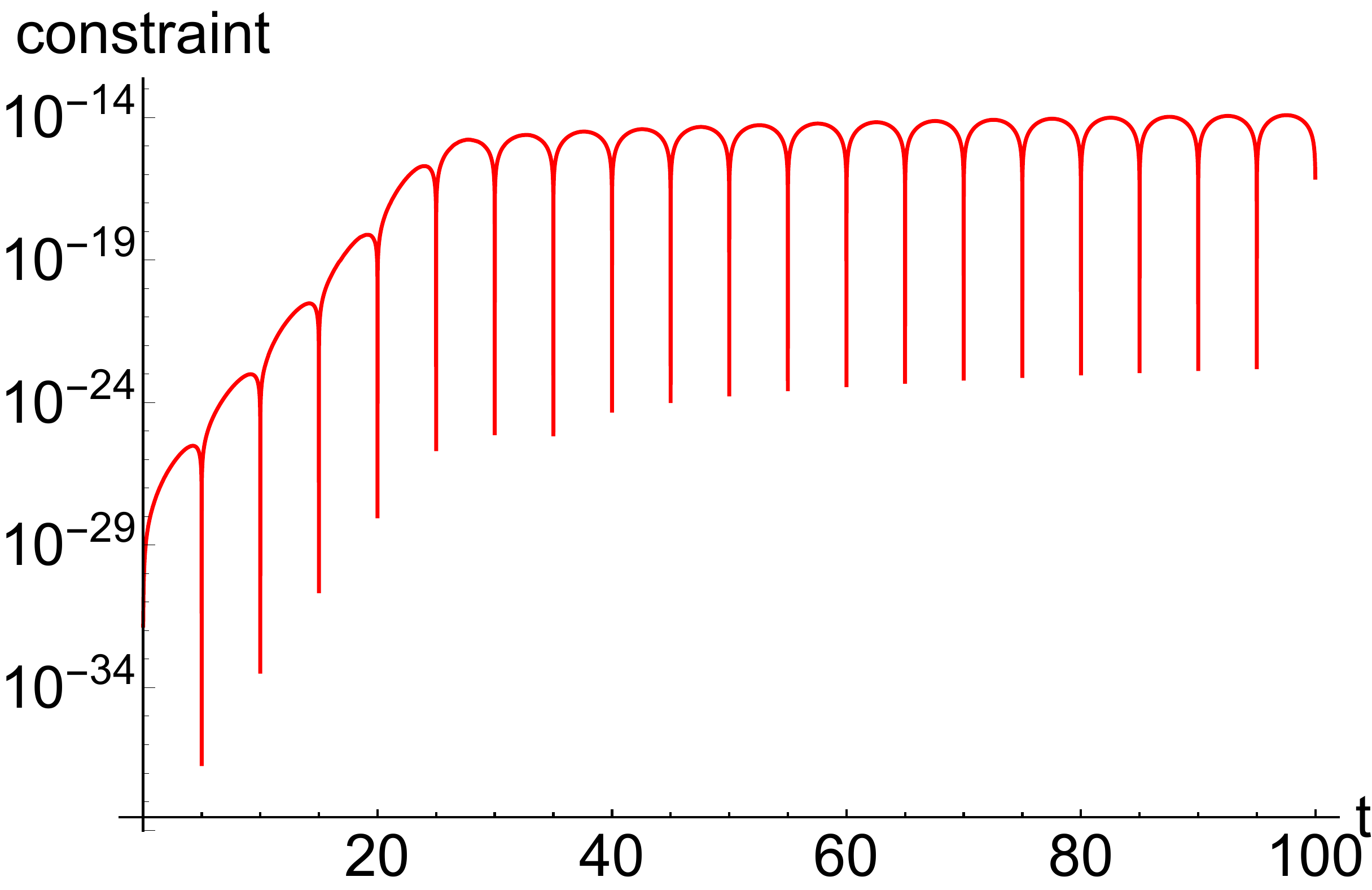}
    \caption{The stress-strain signals, the energy density and the constraint equation \eqref{constraint} for $s_1|_{t=0}=0, \gamma_0=0.1, w_c=2, t_c=25$ for \color{black} $\omega/m=0.32$. \color{black}}
    \label{example}
\end{figure}\\
Following the characteristic feature of the bulk equations of motion in EF coordinates \cite{Chesler:2013lia},  here is the numerical routine to solve the equations of motion \eqref{constraint}-\eqref{eq-dppS}:
\begin{enumerate}
\item We start with a static black hole solution with $\tilde{H}=0, \tilde{S}=s_1(t_0)$ as an initial configuration and choose a strain function $\gamma(t)$.
\item We check the accuracy of numerical calculation, by plugging $\tilde{H}, \tilde{S}$ in the constraint equation \eqref{constraint}. 
\item We use the definition of apparent horizon, $d_+S(u=1)=0$,  as a boundary condition to calculate $\widetilde{d_+S}$ by solving eq. \eqref{eq-dpS}.
\item Then we solve eq. \eqref{eq-dpH} with one boundary condition for $\widetilde{d_+H}$ at asymptotic region, $\widetilde{d_+H}(u=0)=\tfrac{\dot{h_0}\dot{s_0}}{s_0}+\ddot{h_0}$.
\item Now we can solve eq. \eqref{eq-A} to find $\tilde{A}$ with two boundary conditions: the first is $\tilde{A}(u=0)=2(s_1-\dot{s_0})/s_0$ and the second we can find by expanding eq. \eqref{eq-dppS} near the horizon which leads to
\begin{equation}
A =-\frac{1 }{3} \left({d_+H}\right)^2\bigg|_{u=1}.
\end{equation}
\item By using the definition of operator $d_+$  we find the $\dot{\tilde{S}}, \dot{\tilde{H}}$. Then we integrate in time, by employing fourth-order Runge-Kutta method for the first three time steps and then  the fourth order Adams-Bashforth method, to compute $\tilde{H}(u, t_0+\delta t)$ and $\tilde{S}(t_0+\delta t)$ and repeat the same routine from step 2.
\end{enumerate}

To impose the sinusoidal strain, we turn on the amplitude smoothly (in the spirit of \cite{Ammon:2016fru,Grieninger:2017jxz}) as
\begin{equation}
    \gamma(t)= \frac{\gamma_0}{2}\left(1+\tanh\left(\frac{t-t_c}{w_c}\right)\right)\sin\left(2\pi\omega t\right),
\end{equation}
where parameters $t_c, w_c$ control how accurate the initial configuration satisfies the constraint equation \eqref{constraint} and how fast the maximum strain is reached respectively. In fig.\eqref{example} we show the stress, the energy density and we check the constraint equation \eqref{constraint} for a concrete example.
\subsection{More about the nonlinear response}\label{app4}
\begin{figure}
    \centering
    \includegraphics[width=0.7 \linewidth]{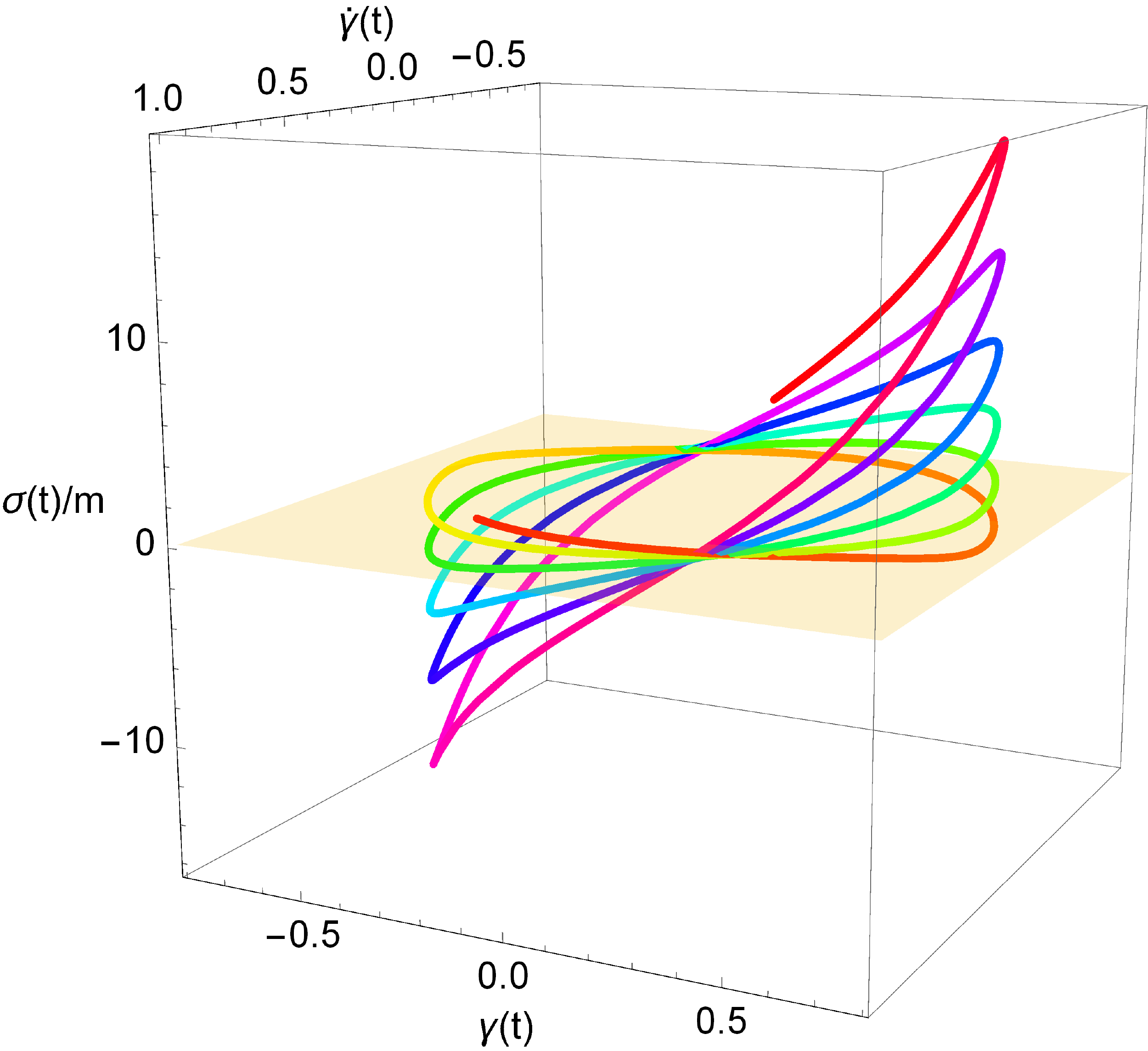}
    
    \vspace{0.4cm}
    
     \includegraphics[width=0.7 \linewidth]{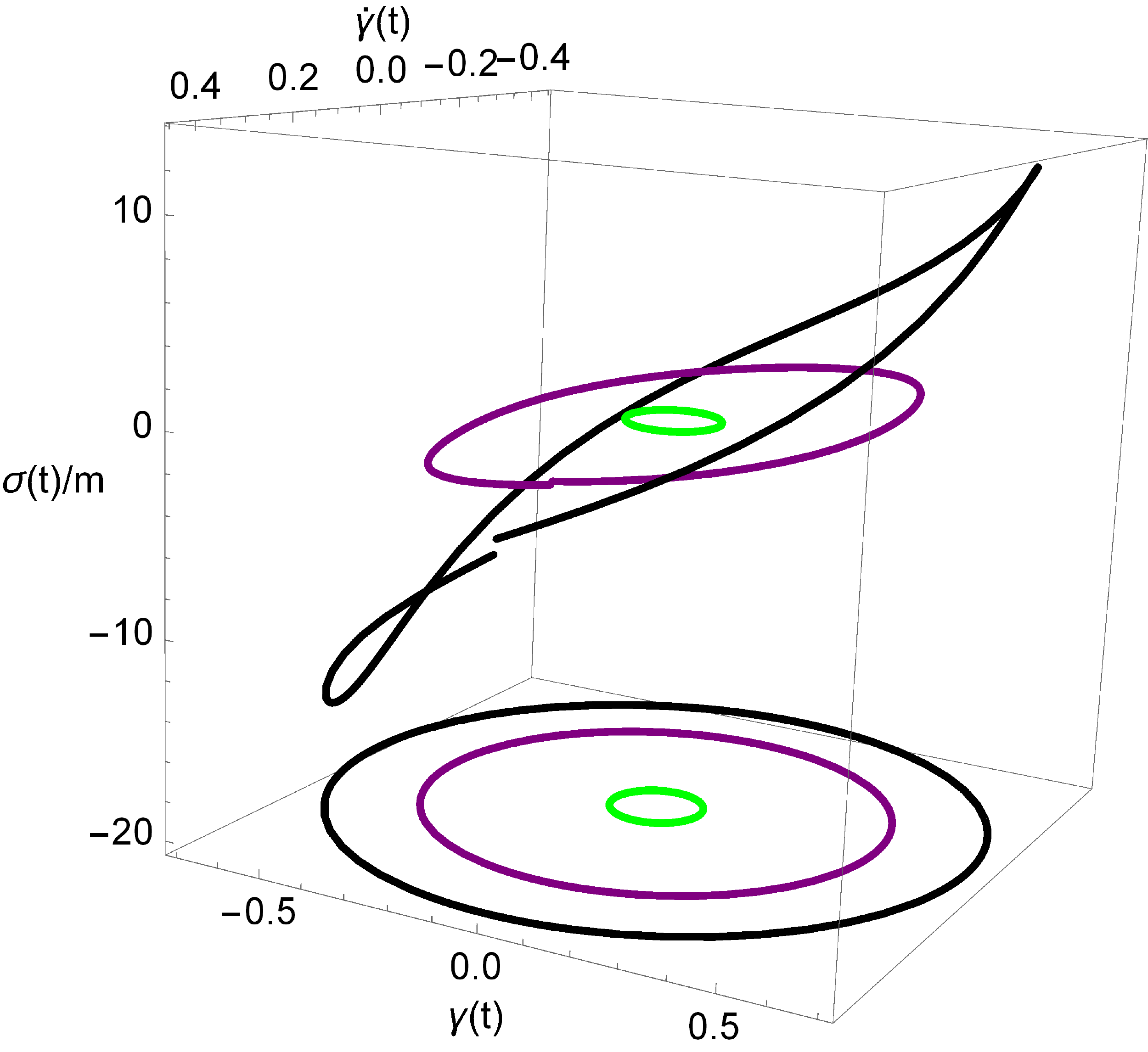}
    \caption{The three dimensional phase space $\{\color{black}\sigma/m\color{black},\gamma,\dot{\gamma}\}$ for an oscillatory strain $\gamma(t)=\gamma_0\,\sin (2\omega t)$. \color{black} We fix $m/T_{in}= 1.81$, $\omega/m= 0.32$\color{black}. In the top panel we fix $\gamma_0=0.7$, while in the bottom we show the results for increasing $\gamma_0=0.1,0.5,0.7$ (from green to black).}
    \label{fig:3qq}
\end{figure}

\begin{figure}[h!]
    \centering
    \includegraphics[width=0.7 \linewidth]{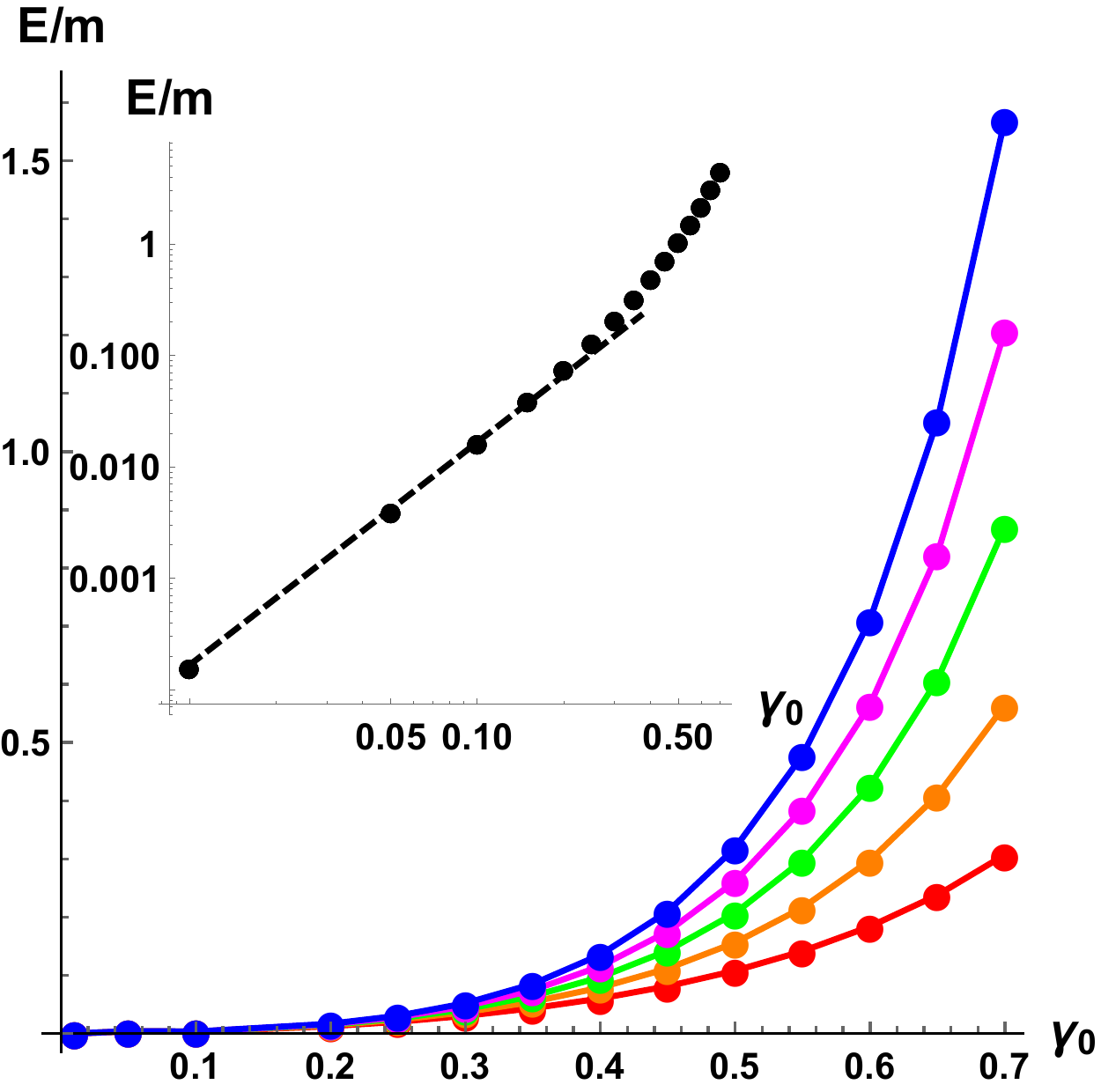}
    \caption{The dissipated energy $\mathrm{E}$ over the first $5$ cycles in function of the strain amplitude $\gamma_0$. The inset shows the average over the first $5$ cycles and the low amplitude scaling $\sim \gamma_0^2$.}
    \label{fig:4mm}
\end{figure}

\begin{figure}[h]
    \centering
    \vspace{0.2cm}
    \includegraphics[width=0.7 \linewidth]{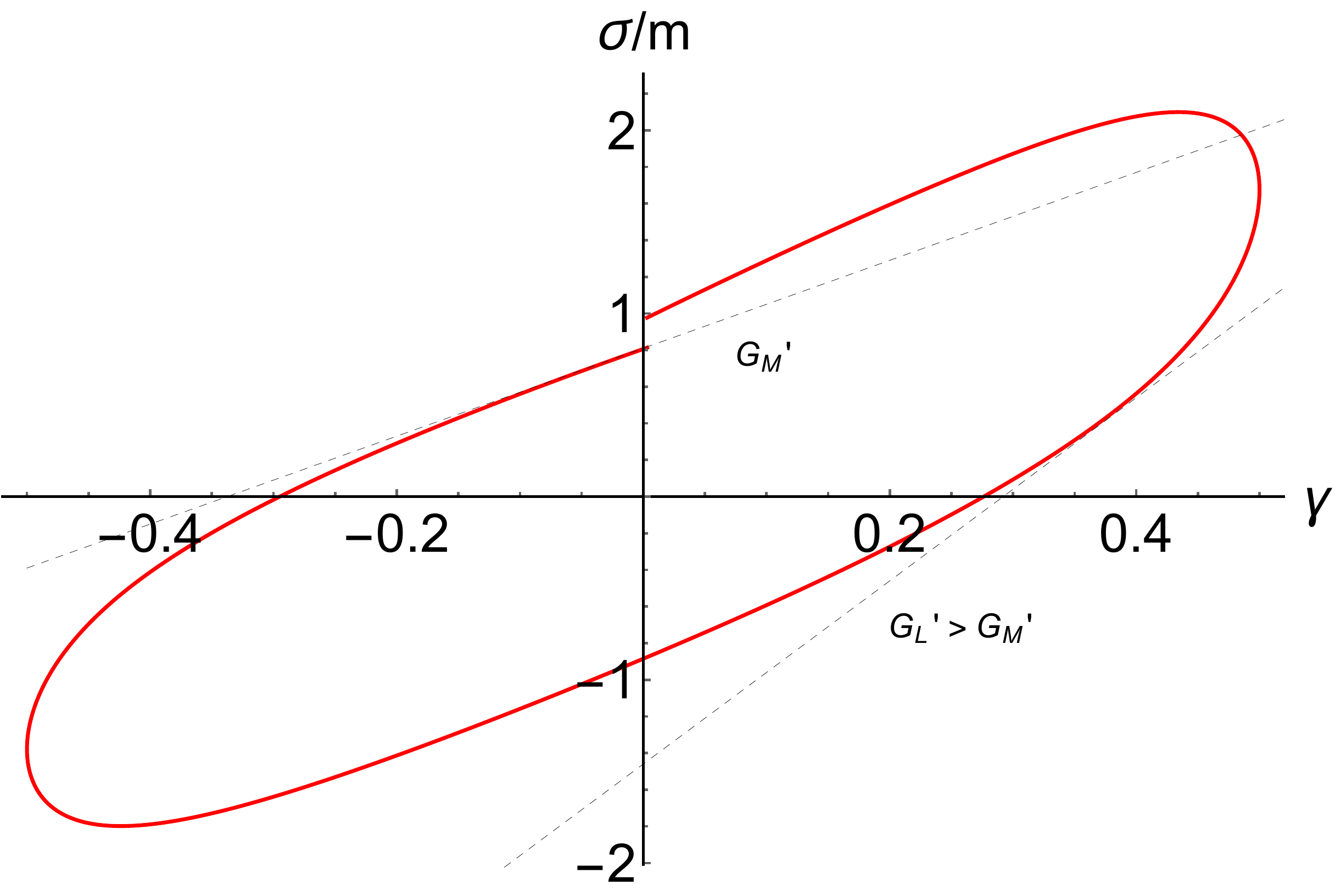}
    
    \vspace{0.4cm}
    
     \includegraphics[width=0.7 \linewidth]{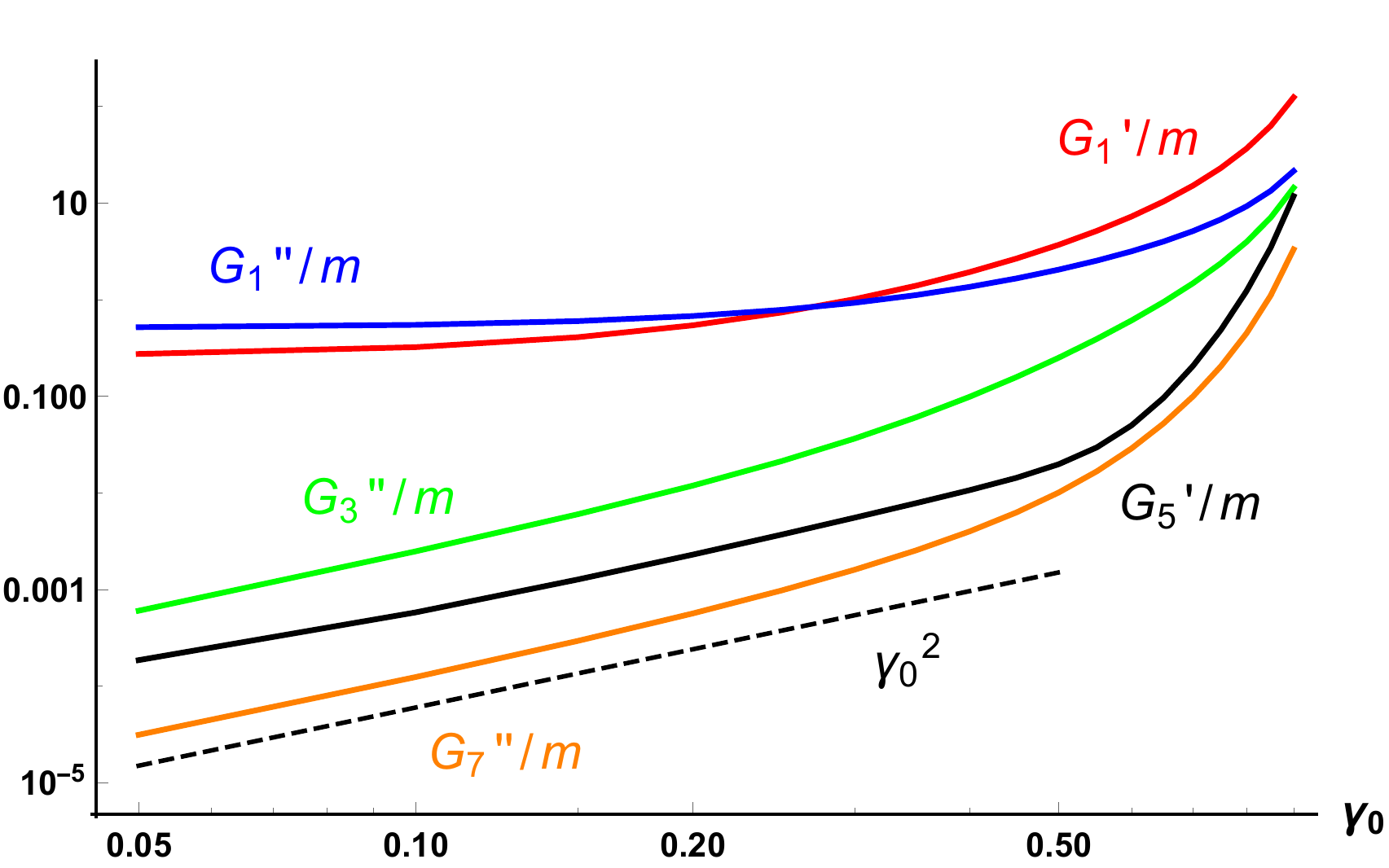}
    \caption{\textbf{Top: }An example of Lissajous figure and the extraction of the small strain and large strain moduli. In this figure $G'_L>G'_M$. This implies $\mathrm{S}>0$ and a \textit{strain stiffening} phenomenon. \textbf{Bottom: }The higher harmonics moduli in function of the strain amplitude $\gamma_0$.}
    \label{fig:5}
\end{figure}

\begin{figure}[b!]
  \centering
  \includegraphics[width=0.7 \linewidth]{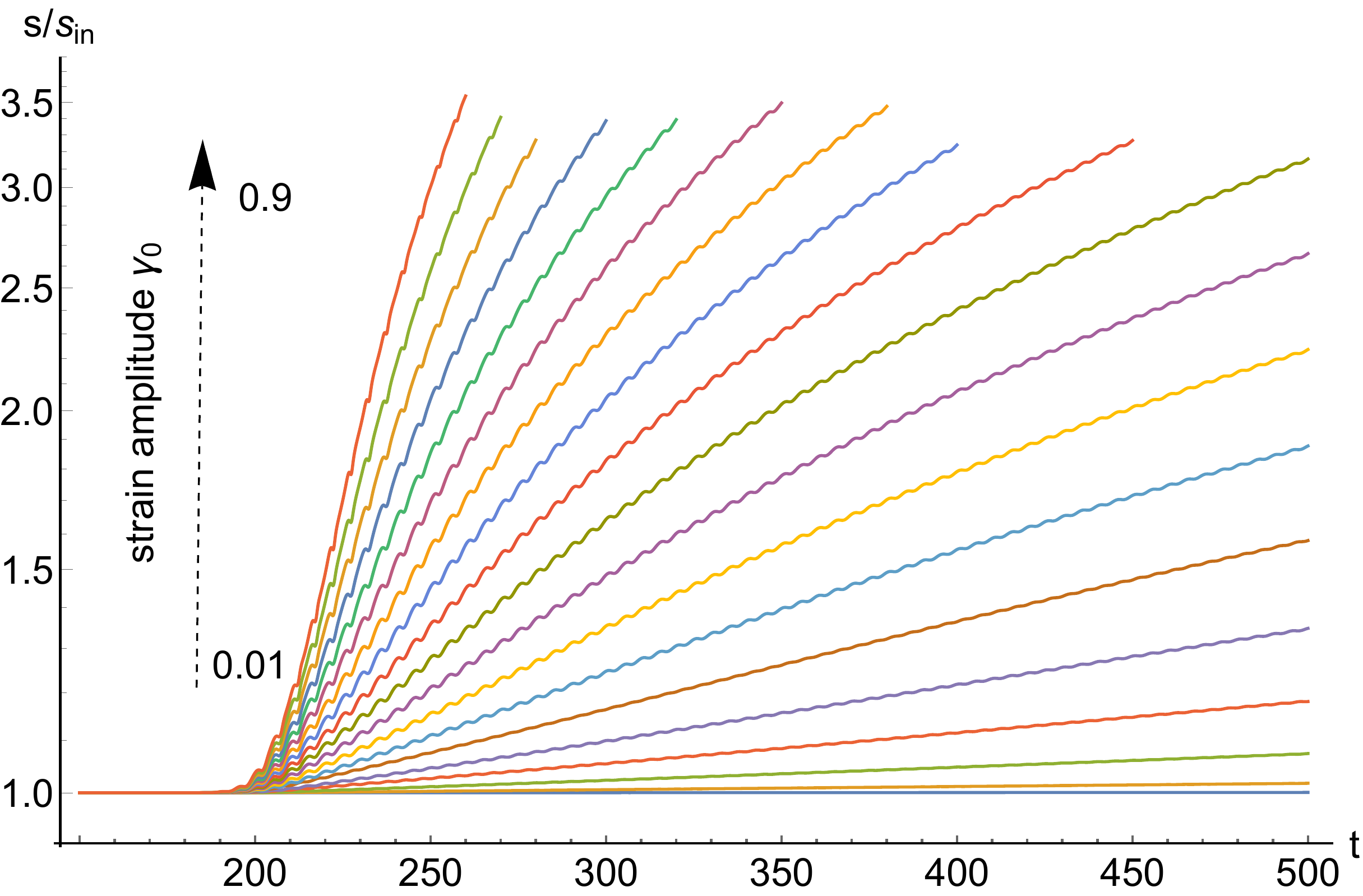}
  
  \vspace{0.4cm}
  
    \includegraphics[width=0.7 \linewidth]{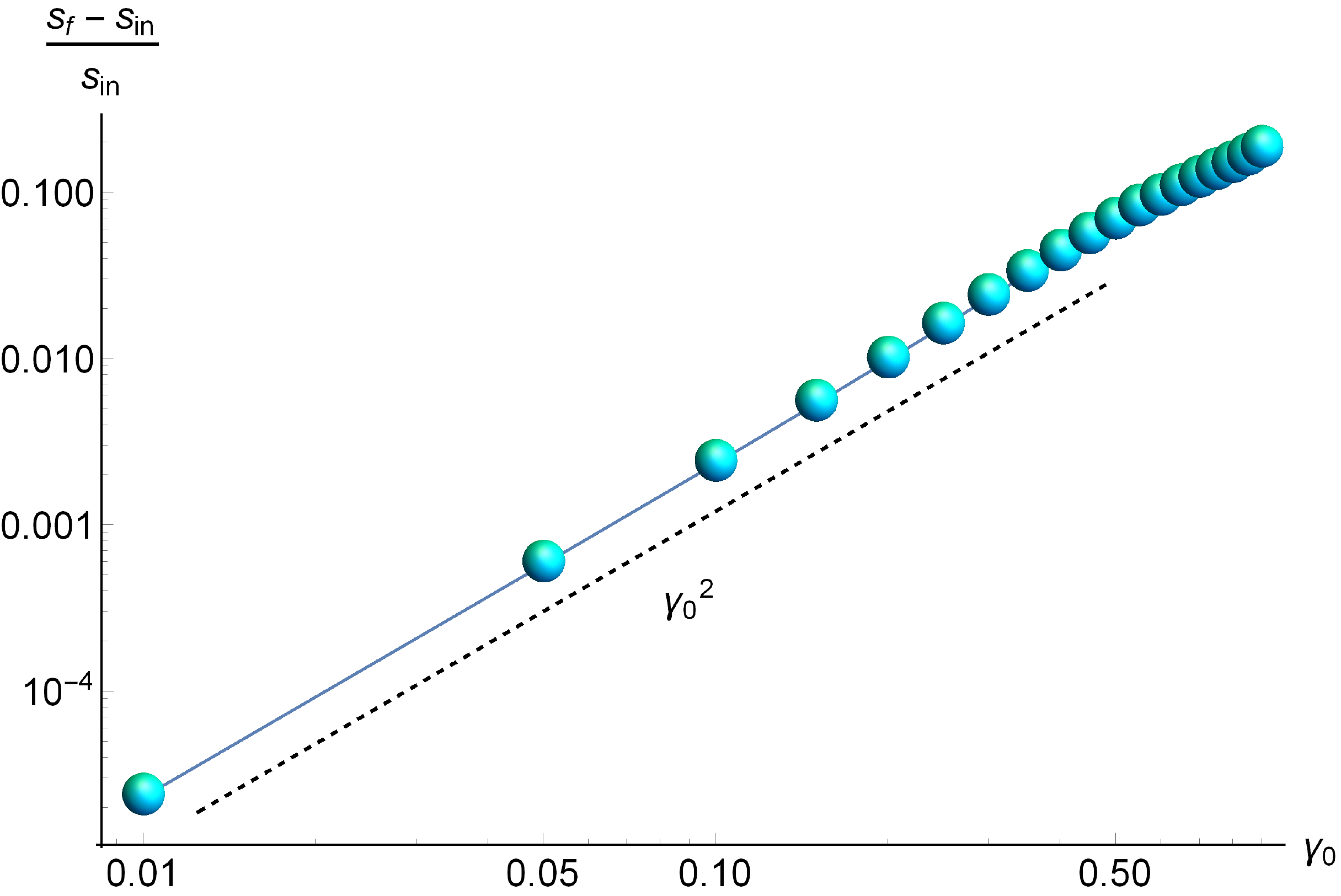}
     \caption{\textbf{Top: }The time dependent entropy density $s\equiv \mathcal{A}/4\pi$ for various strain amplitudes \color{black} normalized by its initial value $s_{in} \equiv s(t=0)$. \color{black} \textbf{Bottom: }The dependence of the percentual entropy increase with respect to the amplitude strain $\gamma_0$ and the low amplitude scaling $\sim \gamma_0^2$.}
    \label{fig:e1}
\end{figure}

In this last section, we give some more details about the nonlinear response. First, we show the shape of the 3D phase space $\{\sigma,\gamma,\dot{\gamma}\}$. In the top panel of fig.\ref{fig:3qq}, we display it for a large value of the strain, within the LAOS regime. The distorted shape of the curves there are indeed the confirmation that nonlinear corrections are important in size.
\\
In the same figure (bottom panel), we emphasize the transition into the nonlinear regime by plotting the 3D curve for small, intermediate and large strain. The difference between the linear green curve and the large amplitude black one is evident.
The Lissajous figures shown in the main text are very good indicators of the viscoelastic behavior and they contain important information about the dissipative mechanism in the system. More specifically, the area of the figure coincides indeed with the dissipated energy over the cycle:
\begin{equation}
    \mathrm{E}\,\equiv\,\int_\mathcal{C} \sigma(\gamma)\, d\gamma
\end{equation}
where $\mathcal{C}$ indicates that the integral is performed over a specific cycle.
We plot the dissipated energy in function of the strain amplitude in fig.\ref{fig:4mm}. We observe that increasing the amplitude of the applied strain the dissipated energy grows. Moreover, at low strain, a scaling regime appears, where $\mathrm{E}\sim \gamma_0^2$.
\\
Finally, we discuss in more detail the nonlinear response in terms of the complex moduli and the entropy production. Given a Lissajous figure, we can define two important quantities:
\begin{equation}
    G'_M\,\equiv\,\frac{d \sigma}{d\gamma}\big|{\gamma=0}\,,\quad G'_L\,\equiv\,\frac{d \sigma}{d\gamma}\big|{\gamma=\gamma_{max}}
\end{equation}
where $G'_M$ is the small strain (tangent) modulus and $G'_L$ the large strain (secant) modulus. A valid indicator of the nonlinear response is the difference between those two values:
\begin{equation}
    \mathrm{S}\,\equiv\,\frac{G'_L\,-\,G'_M}{G'_M}
\end{equation}
Whenever $\mathrm{S}>0$, the nonlinear response exhibits \textit{strain stiffening} -- a larger elastic response at larger applied strains. We generally observe that our system displays this feature independently of the parameters (see top panel of fig.\ref{fig:5} for a concrete choice). This confirms the analysis made in the main text regarding $G'_1$ and $G''_1$.
The behavior of the $n^{th}$ complex moduli is repeated in the bottom panel of fig.\ref{fig:5}. As already shown in the main text, at low strain amplitudes $G'_1,G''_1$ are roughly independent of $\gamma_0$ and they correspond to the linear moduli $G',G''$. At larger strains they start growing signaling the \textit{strain stiffening} phenomenon. On the contrary, the higher harmonics moduli $G_n$ are zero at small strain amplitude -- the response is fully linear. They then grow, rendering the stress response highly nonlinear.\\
To conclude, we investigate the behavior of the entropy density during the nonlinear dynamics. The results are shown in fig. \ref{fig:e1}. We observe that the entropy density grows rapidly with time, as indicated by the growing of the BH horizon area. Moreover, we conclude that the entropy density grows in function of the applied strain amplitude. At least for intermediate values of the amplitude, the growth is very well approximated by a quadratic function $\sim \gamma_0^2$. At this stage, it is not clear to us how to relate this behavior with the strain stiffening that we observe in the nonlinear viscoelastic response.
\balance

\end{document}